\documentclass[journal]{IEEEtran}
\usepackage{graphicx,color}
\usepackage{mathrsfs}
\usepackage{amsmath, setspace}
\usepackage{cite}
\usepackage{amsthm,amssymb,amsfonts,latexsym,bm}
\usepackage{amsfonts}
\usepackage{enumerate}
\usepackage{epstopdf}
\usepackage{subfigure, makecell}

\ifCLASSINFOpdf
\else
\fi
\begin{document}
%
% paper title
% Titles are generally capitalized except for words such as a, an, and, as,
% at, but, by, for, in, nor, of, on, or, the, to and up, which are usually
% not capitalized unless they are the first or last word of the title.
% Linebreaks \\ can be used within to get better formatting as desired.
% Do not put math or special symbols in the title.
\title{Sparsity Learning Based Multiuser Detection in Grant-Free Massive-Device Multiple Access}
%
%
% author names and IEEE memberships
% note positions of commas and nonbreaking spaces ( ~ ) LaTeX will not break
% a structure at a ~ so this keeps an author's name from being broken across
% two lines.
% use \thanks{} to gain access to the first footnote area
% a separate \thanks must be used for each paragraph as LaTeX2e's \thanks
% was not built to handle multiple paragraphs
%

\author{Tian~Ding,
	Xiaojun~Yuan,~\IEEEmembership{Senior Member,~IEEE,}
	and~Soung~Chang~Liew,~\IEEEmembership{Fellow,~IEEE}
	
	\thanks{This work was presented in part at the IEEE GLOBECOM, Abu Dhabi, UAE, Dec. 2018 \cite{ref_Conf}.}
	
	\thanks{T. Ding and S. C. Liew are with the Department of Information Engineering, the Chinese University of Hong Kong. X. Yuan is with the Center for Intelligent Networking and Communications (CINC), the University of Electronic Science and Technology of China, Chengdu, China.}
	
	\thanks{The work of S. C. Liew and T. Ding was supported in part by the General Research Funds (Project No. 14200417) established under the University Grant Committee of the Hong Kong Special Administrative Region, China. The work of Xiaojun Yuan was supported in part by The Key Area Research and Development Program of Guangdong Province, China (Project No. 2018B010114001).}
}
\maketitle

% As a general rule, do not put math, special symbols or citations
% in the abstract or keywords.
\begin{abstract}
In this work, we study the multiuser detection (MUD) problem for a grant-free massive-device multiple access (MaDMA) system, where a large number of single-antenna user devices transmit sporadic data to a multi-antenna base station (BS). Specifically, we put forth two MUD schemes, termed random sparsity learning multiuser detection (RSL-MUD) and structured sparsity learning multiuser detection (SSL-MUD) for the time-slotted and non-time-slotted grant-free MaDMA systems, respectively. In RSL-MUD, active users generate and transmit data packets with random sparsity. In SSL-MUD, we introduce a sliding-window-based detection framework, and the user signals in each observation window naturally exhibit structured sparsity. We show that by exploiting the sparsity embedded in the user signals, we can recover the user activity state, the channel, and the user data in a single phase, without using pilot signals for channel estimation and/or active user identification. To this end, we develop a message-passing-based statistical inference framework for the BS to blindly detect the user data without any prior knowledge of the identities and the channel state information (CSI) of active users. Simulation results show that our RSL-MUD and SSL-MUD schemes significantly outperform their counterpart schemes in both reducing the transmission overhead and improving the error behavior of the system.
\end{abstract}

% Note that keywords are not normally used for peerreview papers.

% For peer review papers, you can put extra information on the cover
% page as needed:
% \ifCLASSOPTIONpeerreview
% \begin{center} \bfseries EDICS Category: 3-BBND \end{center}
% \fi
%
% For peerreview papers, this IEEEtran command inserts a page break and
% creates the second title. It will be ignored for other modes.
%\IEEEpeerreviewmaketitle

\section{Introduction}
\IEEEPARstart{M}{assive-device} multiple access (MaDMA) is an emerging research topic for next generation wireless communication. In a typical massive-device communication network, hundreds or even thousands of user devices are associated with a single cellular base station (BS), with only a small fraction of them being actively and wanting to communicate with the BS at a time. The BS is required to dynamically identify the active users and reliably receive their messages. Such a model arises in many practical scenarios, e.g., wireless sensor networks \cite{ref_WSN} and Internet of Things (IoT) \cite{ref_IoT}.

Two considerations in the design of a MaDMA system are as follows: First, user devices are usually kept in a sleep mode to save energy and are activated only when certain external events occur. Therefore, the data traffic of each user node is generated sporadically. Typically only a small fraction of the devices are activated for transmission at any time. Second, the system overhead for identifying the active devices and coordinating their transmissions should be kept to a minimum given the large number of user devices.

\begin{figure}
	\center
	\includegraphics[width=8cm]{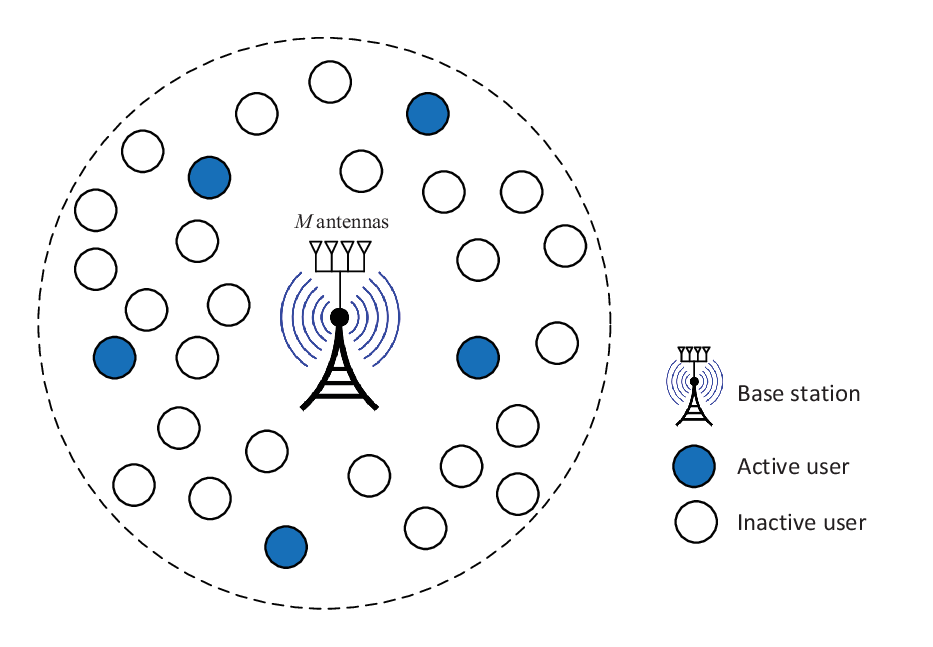}\\
	\caption{System model of the massive-device network.}\label{fig_system_model}
\end{figure}

In conventional cellular networks, user transmissions are scheduled into different time or frequency slots via a contention procedure over a separate control channel. This idea has been extended to massive-device networks. For example, \cite{ref_contention1,ref_contention2,ref_contention3} studied a contention-based multiple access protocol, where each active user randomly picks a signature sequence and sends it to the BS. If the selected preamble is not used by any other users, the active user can set up a connection with the BS. However, contention-based protocols suffer from potential collisions, and the contention phase may introduce excessive overhead for control signalling due to the large number of potential users in a MaDMA system. As such, grant-free protocols, where user devices initiate data transmission without any handshaking process with the BS, are more desirable in massive-device networks \cite{ref_NOMA}.

Multiuser detection (MUD) plays an essential role in the design of a grant-free MaDMA system\footnote{
	In prior works, grant-free MaDMA was also termed grant-free non-orthogonal multiple access (NOMA) \cite{ref_CSMUD_CDMA3, ref_CSMUD_CDMA4, ref_CSMUD_CDMA6, ref_CSMUD_CDMA7, ref_CSMUD_other1, ref_CSMUD_other4, ref_CSMUD_noCSI7}, massive machine-type communication \cite{ref_CSMUD_CDMA2, ref_CSMUD_CDMA5, ref_CSMUD_noCSI1, ref_CSMUD_noCSI2}, or massive connectivity \cite{ref_CSMUD_noCSI4, ref_CSMUD_noCSI5, ref_CSMUD_noCSI6}.} because, without coordination, simultaneous transmissions by multiple devices are possible. In particular, BS needs to detect multiple user signals. Furthermore, it needs to do so without any prior knowledge of the identities of active users. The MUD problem involves three key components: active user identification, channel estimation, and data decoding. In existing works \cite{ref_CSMUD_CDMA1, ref_CSMUD_CDMA2, ref_CSMUD_CDMA3, ref_CSMUD_CDMA4, ref_CSMUD_CDMA5,ref_CSMUD_CDMA6,ref_CSMUD_CDMA7, ref_CSMUD_other1, ref_CSMUD_other2, ref_CSMUD_other3, ref_CSMUD_other4,ref_CSMUD_noCSI1, ref_CSMUD_noCSI2, ref_CSMUD_noCSI3, ref_CSMUD_noCSI4, ref_CSMUD_noCSI5, ref_CSMUD_noCSI6, ref_CSMUD_noCSI7}, these three components are actualized in two separate phases: Either a channel estimation phase is followed by a joint active user identification and data decoding phase, or a joint channel estimation and active user identification phase takes place before the data decoding.

Compressed sensing (CS) techniques \cite{ref_CS,ref_model_CS} have been extensively used in two-phase MUD schemes for grant-free MaDMA. Specifically, the authors in \cite{ref_CSMUD_CDMA1, ref_CSMUD_CDMA2, ref_CSMUD_CDMA3, ref_CSMUD_CDMA4, ref_CSMUD_CDMA5, ref_CSMUD_CDMA6, ref_CSMUD_CDMA7, ref_CSMUD_other1, ref_CSMUD_other2, ref_CSMUD_other3, ref_CSMUD_other4} assumed perfect channel state information (CSI) at the BS, and then cast the joint active user identification and data decoding as a CS problem. In \cite{ref_CSMUD_noCSI1, ref_CSMUD_noCSI2, ref_CSMUD_noCSI3, ref_CSMUD_noCSI4, ref_CSMUD_noCSI5, ref_CSMUD_noCSI6, ref_CSMUD_noCSI7}, the authors considered a more practical scenario where the CSI is not available at the BS, and the BS is equipped with multiple antennas to achieve spatial multiplexing. Pilot signals are transmitted by users, and the BS exploits CS techniques to conduct joint channel estimation and active user identification based on the pilot signals. Then, the BS decodes the user data based on the estimated CSI of active users.

Potentially, integrating the three components into a single phase can further reduce the transmission overhead and thereby increases the spectral efficiency of the system. The corresponding MUD problem, however, will be more challenging, since the BS needs to identify the active users, estimate the channel, and decode the user data based on only a single phase of the data signals (no pilots, no preambles, etc.). Furthermore, the existing works mostly focused on time-slotted transmission where users can change their activity state only at the beginning of each time slot. In time-slotted transmission, time slot alignment among the massive number of devices will cause significant overhead, and any device that fails to align its time slots properly may screw up the whole MUD process. In that sense, non-time-slotted transmission is more desirable for grant-free MaDMA to minimize the user coordination overhead and to increase reliability. However, non-time-slotted transmission generally renders the active user identification, channel estimation, and data decoding even more challenging tasks.

This paper studies the MUD problem for both time-slotted and non-time-slotted grant-free MaDMA systems. We assume that the BS is equipped with multiple receive antennas and each active single-antenna user sends a sparse signal to the BS. This implies that the signal of each active user received by the BS is a rank-one matrix that can be factorized as the outer product of a channel vector and a sparse signal vector. We show that by appropriately exploiting channel randomness and user signal sparsity, the BS is able to jointly estimate the channels and decode the signals from active users based on efficient matrix factorization algorithms. User identity can then be recovered by extracting the pre-inserted user label in each decoded data packet. This implies that active user identification, channel estimation, and data decoding can be accomplished in a single phase, which significantly reduces transmission overhead and improves system efficiency. This MUD approach falls into the realm of blind multiple input multiple output (MIMO) detection, since unlike the existing works \cite{ref_CSMUD_CDMA1, ref_CSMUD_CDMA2, ref_CSMUD_CDMA3, ref_CSMUD_CDMA4, ref_CSMUD_CDMA5,ref_CSMUD_CDMA6, ref_CSMUD_CDMA7, ref_CSMUD_other1, ref_CSMUD_other2, ref_CSMUD_other3, ref_CSMUD_other4, ref_CSMUD_noCSI1, ref_CSMUD_noCSI2, ref_CSMUD_noCSI3, ref_CSMUD_noCSI4, ref_CSMUD_noCSI5, ref_CSMUD_noCSI6, ref_CSMUD_noCSI7}, no pilot signals are needed for channel estimation and/or active user identification.

To be specific, we develop two different MUD schemes, termed random sparsity learning MUD (RSL-MUD) and structured sparsity learning MUD (SSL-MUD) for the time-slotted and non-time-slotted grant-free MaDMA systems, respectively.\footnote{The term ``sparsity learning" is similar to the term ``sparse Bayesian learning" \cite{ref_sparse_bayesian1,ref_sparse_bayesian2,ref_sparse_bayesian3}, but they have different meanings. Sparse Bayesian learning is to use the Bayesian framework to reconstruct a sparse vector $\mathbf{x}$ from a noisy measurement vector $\mathbf{y} = \mathbf{\Phi x} + \mathbf{w}$. The sensing matrix $\mathbf{\Phi}$ is assumed to be known, and the problem is equivalent to finding solutions to an underdetermined linear system. However, in this paper, sparsity learning refers to sparse matrix factorization, i.e., to simultaneously recover a random channel matrix $\mathbf{H}$ and a sparse signal matrix $\mathbf{X}$ from a noisy measurement matrix $\mathbf{Y} = \mathbf{HX}+\mathbf{W}$. The sparse matrix factorization problem is bilinear, hence more challenging to solve.} In the time-slotted RSL-MUD scheme, users are randomly activated and transmit sparse signals synchronized at a packet level, where the zero positions in a user packet are independently chosen. The blind MIMO detection problem is then cast as a dictionary learning problem \cite{ref_dictionary_learning}, and can be solved by the bilinear generalized approximate message passing (BiG-AMP) algorithm \cite{ref_BiGAMP}. One reference symbol for each user packet is used to eliminate the phase ambiguity inherent in blind MIMO detection.

In the non-time-slotted SSL-MUD scheme, users are allowed to initiate packet transmission at any time, and hence the BS cannot conduct signal detection in a time-slotted manner. Instead, we introduce a sliding-window-based detection framework, where the BS successively detects and decodes the user packets over a moving time window. A key observation is that user devices in a non-time-slotted system usually transmit short packets in an intermittent manner. This implies that the user signals in a relatively large time window naturally exhibit \textit{structured sparsity}, i.e., the signals of a user are mostly zeros while the non-zeros are clustered together to form a packet with a small number of symbols. For blind MIMO detection at the BS, we put forth a message-passing-based algorithm, termed turbo bilinear generalized approximate message passing (Turbo-BiG-AMP). Turbo-BiG-AMP partitions the overall graphical model of the detection problem into two subgraphs: one modelling the bilinear constraints of the channel model, and the other modelling the structured sparsity of the user signals. The inference on the two subgraphs iterates until convergence, following the idea of turbo compressed sensing \cite{ref_Turbo1,ref_Turbo2,ref_Turbo3}.

The main contributions of this paper are summarised as follows. To the best of our knowledge, this is the first attempt to address the MUD problem in MaDMA systems by incorporating active user identification, channel estimation, and data decoding in a single phase. To this end, we develop a message-passing-based statistical inference framework to systematically and efficiently solve the blind MIMO MUD problem in both time-slotted and non-time-slotted MaDMA systems. We show that our RSL-MUD and SSL-MUD schemes significantly outperform their counterpart schemes in the literature in both reducing the transmission overhead and improving the error behavior of the system.

\subsection*{Paper Organization and Notation}
In Section II, we present the system models of the time-slotted and non-time-slotted grant-free MaDMA systems. Sections III and IV develop the RSL-MUD and SSL-MUD schemes, respectively. The Turbo-BiG-AMP algorithm for the blind MIMO detection problem in SSL-MUD is also put forth in Section IV. Section V presents simulation results for both RSL-MUD and SSL-MUD. Finally, in Section VI, we conclude the paper.

Throughout this paper, we use san-serif font, e.g., $\mathsf{x}$, for random variables and serif font, e.g. $x$, for other variables. We use bold upper and lower case letters for matrices and column vectors, respectively. $\mathbb{C}^{m\times n}$ denotes the $m \times n$ dimensional complex space. $\mathbf{0}$ and $\mathbf{I}_{n}$ represent
the all-zero matrix and the $n$-dimensional identity matrix,
respectively. We then use $p_{\mathsf{x}}(x)$ to denote the pdf of random variable $\mathsf{x}$, and $\mathcal{CN}(x;\hat{x},\nu^x)$ to denote the complex Gaussian pdf for a scalar random variable with mean $\hat{x}$ and
variance $\nu^x$. For a matrix $\mathbf{X}$, we use $x_{i,j}$ to denote the entry in the $i$-th row and $j$-th column. For a vector $\mathbf{x}$, we use $x(i)$ to denote the $i$-th entry. The summary of key notations in this paper is presented in Table \ref{tab_notations} for convenience.

\begin{table}
	\caption{Summary of Key Notations}
	\small
	\label{tab_notations}
	\begin{center}
		\begin{tabular}{ |c|l| }
			\Xhline{1.2pt}
			\multicolumn{2}{|c|}{System model}\\
			\Xhline{1.2pt}
			$K$ & Number of user devices \\
			\hline
			$M$ & Number of antennas at the BS \\
			\hline
			$\mathcal{I}_U$ & Index set of user devices, i.e., $\{1,2,\cdots,U\}$\\
			\Xhline{1.2pt}
			\multicolumn{2}{|c|}{Time-slotted RSL-MUD}\\
			\Xhline{1.2pt}
			$T$ & Length of time slot\\
			\hline
			$\alpha_i$ & Activity indicator of user $i$\\
			\hline
			$\mathcal{N}$ & Set of active users \\
			\hline
			$N$ & Number of active users, i.e., $|\mathcal{N}|$ \\
			\Xhline{1.2pt}
			\multicolumn{2}{|c|}{Non-time-slotted SSL-MUD}\\
			\Xhline{1.2pt}
			$L$ & Length of user packet \\
			\hline
			$t^{(j)}_i$ &  Start time of the $j$-th packet of user $i$ \\
			\hline
			$t_k$ & Start time of the $k$-th observation window\\
			\hline
			$T'$ & Length of observation window\\
			\hline
			$\Delta t$ & Step size of sliding window\\
			\hline
			$N_{t_k}$ & Number of active packets in window $[t_k, t_k+T')$\\
			\Xhline{1.2pt}
		\end{tabular}
	\end{center}
\end{table}

\section{System Model}
This section presents the MUD problem for the grant-free MaDMA systems. Consider a single-cell network where $U$ user nodes, indexed by $ \mathcal{I}_{U} \triangleq \left\{1,2,\cdots,U\right\}$, are associate with a multi-antenna BS, as shown in Fig. \ref{fig_system_model}. Each user is equipped with one antenna and the BS is equipped with $M$ antennas. We assume that the number of users is much greater than the number of antennas at the BS, i.e., $U \gg M$. Moreover, each user generates sporadic data traffic. Thus, although the number of users is large, only a small fraction of them are active at any one time.

In this work, we study two types of grant-free MaDMA: time-slotted grant-free MaDMA and non-time-slotted grant-free MaDMA.

\subsection{Time-Slotted Grant-Free MaDMA}
In time-slotted grant-free MaDMA, the uplink transmission occurs in multiple time-slots, each of $T$ symbol durations. Packet-level synchronization is assumed. Users can only begin the transmission of a packet at the beginning of a time slot and the transmission of the packet must end by the end of the same time slot. In each time slot, only a subset of users are actively transmitting.

For an arbitrary time slot, let $\alpha_i$ be the indicator of user $i$'s activity state, i.e.,
\begin{equation}
\alpha_i = \begin{cases}
1,&\text{user $i$ is active}\\
0,&\text{otherwise}.
\end{cases}
\end{equation}
The activity states $\{\alpha_i\}$ for different $i$ are assumed to be independently and identically distributed (i.i.d.) with $\Pr \{\alpha_i = 1\} = p_1$. Define the set of active users as
\begin{equation}
\mathcal{N} \triangleq \{i:i\in \mathcal{I}_U, \alpha_i = 1\}.
\end{equation}
During the time slot, each active user $i$, $i\in \mathcal{N}$, transmits a data packet $\mathbf{c}_{i} \in \mathbb{C}^{T \times 1}$ to the BS. Each data packet carries not only the desired data message to the BS, but also the identity of the transmitting user. For all active users, we assume an equal power constraint, given by
\begin{equation}
\label{eq_power_TS}
\frac{1}{T}\mathrm{E}\left[\textbf{c}_i^H \textbf{c}_i\right] \leq P, \quad \!\!\forall i \in \mathcal{N}.
\end{equation}

We adopt a block-fading model: the channels of all active users incur independent quasi-static flat fading with coherence time no less than $T$, and hence remain constant during each time slot. Specifically, the complex channel vector from user $i$ to the $M$ antennas of the BS is modeled as
\begin{equation}
\label{eq_channel_model_TS}
\mathbf{h}_i = \sqrt{\beta_i}\mathbf{g}_i \in \mathbb{C}^{M\times 1}
\end{equation}
where $\mathbf{g}_i \sim \mathcal{CN}\left(\mathbf{0}, \mathbf{I}_M\right)$ consists of Rayleigh fading components, and $\beta_i$ is the path-loss and shadowing component, which depends on user $i$'s location and remains the same for all packets transmitted by user $i$. Then, the received signal at the BS can be written as
\begin{equation}
\label{eq_received_signal_TS}
\mathbf{Y} = \sum_{i\in \mathcal{N}}\mathbf{h}_i\mathbf{c}_i^T + \mathbf{W} \in \mathbb{C}^{M \times T}
\end{equation}
where $\mathbf{W} \in \mathbb{C}^{M \times T}$ is the additive white Gaussian noise (AWGN) matrix with the entries generated i.i.d. from $\mathcal{CN}\left(0, \sigma^2\right)$.

The objective of the BS in each time slot is to identify all the active users in $\mathcal{N}$ and to decode all the transmitted packets $\{\mathbf{c}_i : i\in \mathcal{N}\}$ based on the received signal \eqref{eq_received_signal_TS}, without knowing the user activity state $\{\alpha_i\}$ and the CSI $\{\mathbf{h}_i\}$.

\subsection{Non-Time-Slotted Grant-Free MaDMA}
\begin{figure}
	\center
	\includegraphics[width=8cm]{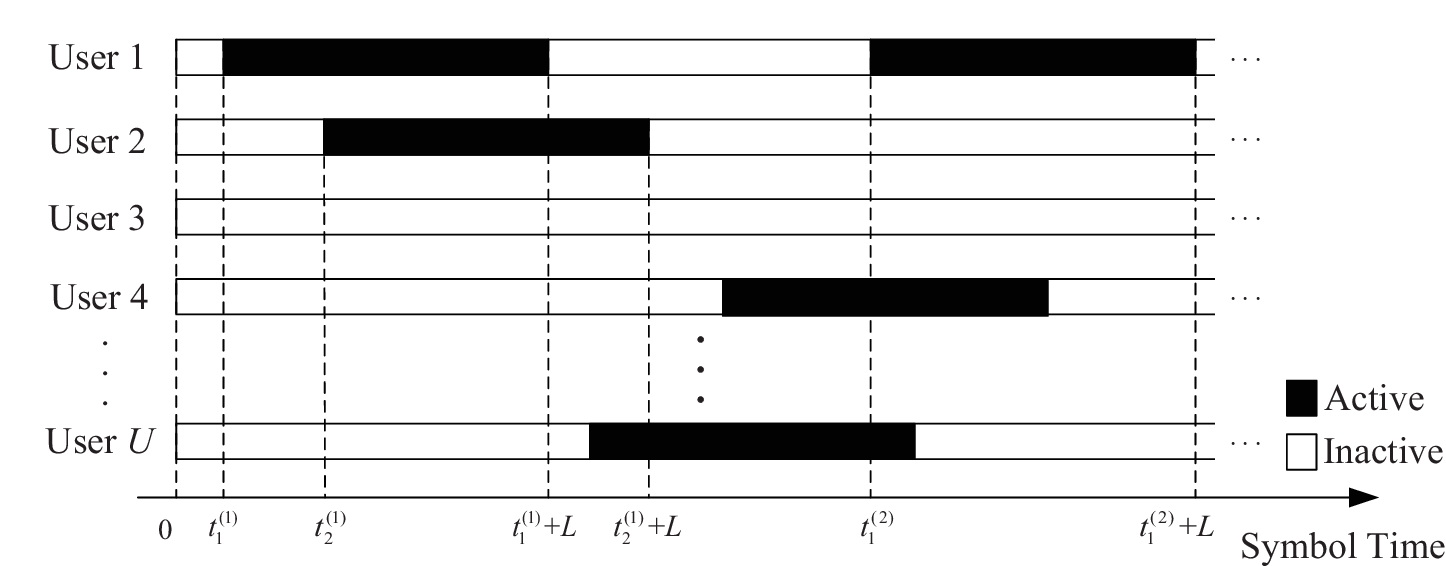}\\
	\caption{The activity states of users over a range of symbol intervals.}\label{fig_activity_state}
\end{figure}

In non-time-slotted grant-free MaDMA, symbol-level synchronization is assumed but not packet-level synchronization. In particular, users are allowed to initiate packet transmission at the beginning of any symbol interval. Whenever a user has a message to transmit, it generates a data packet that carries the message and the identity of the transmitting user. Once activated, the user transmits the packet to the BS in $L$ consecutive symbol intervals, as shown in Fig. \ref{fig_activity_state}.

Let $\mathbf{c}^{(j)}_i \in \mathbb{C}^{L\times 1}$ be the $j$-th packet of user $i$, $s_{i,t}$ be the signal transmitted by user $i$ at the $t$-th symbol interval, and $t^{(j)}_{i}$ be the symbol interval at which user $i$ starts to transmit its $j$-th packet. Then we have
\begin{equation}
s_{i,t^{(j)}_i+k-1} = c^{(j)}_i(k),\quad k=1,2,\cdots,L,\quad\!\! \forall i\in \mathcal{I}_U, \quad \!\! j>0.
\end{equation}
As in time-slotted grant-free MaDMA, users are assumed to have an equal power constraint $\mathrm{E}\left[(\mathbf{c}^{(j)}_i)^H\mathbf{c}^{(j)}_i\right]/L\leq P$. Noting that an inactive user can be seen as an active user transmitting zeros, we have
\begin{equation}
s_{i,t} =
\begin{cases}
c^{(j)}_i(t - t^{(j)}_i + 1),& \text{there exists $j$ such that }\\ &\quad \quad \quad t^{(j)}_i \!\leq\!  t \!<\! t^{(j)}_i + L\\
0,&\text{otherwise}.
\end{cases}
\end{equation}

We adopt the same channel model as in time-slotted grant-free MaDMA. Specifically, for packet $\mathbf{c}^{(j)}_i$, the complex channel vector from user $i$ to the $M$ antennas of the BS is represented by
\begin{equation}
\label{eq_channel_model_NTS}
\mathbf{h}_i^{(j)} = \sqrt{\beta_i}\mathbf{g}^{(j)}_i \in \mathbb{C}^{M\times 1}
\end{equation}
which remains constant during the transmission of $\mathbf{c}^{(j)}_i$. In \eqref{eq_channel_model_NTS}, $\mathbf{g}_i^{(j)} \sim \mathcal{CN}\left(\mathbf{0}, \mathbf{I}_M\right)$ is the vector of Rayleigh fading components, and $\beta_i$ is the path-loss and shadowing component of user $i$. Further, for each user, we assume that there is a sufficient guard period between two consecutive packets, so that the channel vector $\mathbf{h}_i^{(j)}$ varies independently for different packets of the same user.

At the $t$-th symbol interval, the received signal at the BS is represented by
\begin{equation}
\mathbf{y}_t = \sum_{i=1}^{U}\mathbf{h}_{i,t}s_{i,t}+\mathbf{w}_t \in \mathbb{C}^{M\times 1}
\end{equation}
where
\begin{equation}
\mathbf{h}_{i,t} =
\begin{cases}
\mathbf{h}^{(j)}_i, &\text{there exists $j$ such that } t^{(j)}_i \!\leq\!  t \!<\! t^{(j)}_i + L\\
\mathbf{0},&\text{otherwise}
\end{cases}
\end{equation}
and $\mathbf{w}_t$ is the additive white Gaussian noise (AWGN) vector generated from $\mathcal{CN}\left(\mathbf{0}, \sigma^2 \mathbf{I}_M\right)$.

The objective of the BS is to successively detect and decode the transmitted packets $\{\mathbf{c}^{(j)}_i:i\in \mathcal{I}_U,j\geq 0\}$ based on the received signal $\left\{\mathbf{y}_t : t\geq 0 \right\}$, without knowing the transmission times $\{t^{(j)}_{i}\}$ of the user packets and the corresponding CSI $\{\mathbf{h}_i^{(j)}\}$.

\section{RSL-MUD for Time-Slotted Grant-Free MaDMA}
\label{section_RSLMUD}
In this section, we develop the RSL-MUD scheme for the time-slotted grant-free MaDMA. Inspired by \cite{ref_SCMA}, in RSL-MUD active users in $\mathcal{N}$ transmit data packets with random sparsity, i.e., many zero symbols are randomly and independently placed in the transmitted packets. Specifically, let $\mathcal{X}$ be a modulation constellation with $0 \not\in \mathcal{X}$. Each active user generates a data packet with symbols i.i.d. drawn from distribution
\begin{equation}
\label{eq_symbol_distribution_TS}
\Pr\{c_i(t) \!=\! x\}\! =\! \begin{cases}
1\!-\!\gamma, &\! x \!=\! 0 \\
\frac{\gamma}{|\mathcal{X}|}, & \! x \!\in\! \mathcal{X}
\end{cases}, \quad\!\! t\!=\!1,2,\cdots,T, \quad\!\!i \!\in\! \mathcal{N}.
\end{equation}
where $\gamma  \in  (0,1)$ is referred to as \textit{sparsity level}. That is, each symbol is zero with probability $1 - \gamma$, and otherwise uniformly distributed among $\mathcal{X}$. As such, the received signal at the BS can be factorized into a channel matrix and a sparse signal matrix. By exploiting the random sparsity of the signal matrix, the BS is able to jointly estimate the channel and signal matrices, and then recover all the user packets.

\subsection{Sparse Packet Generation}
\begin{figure}
	\center
	\includegraphics[width=8cm]{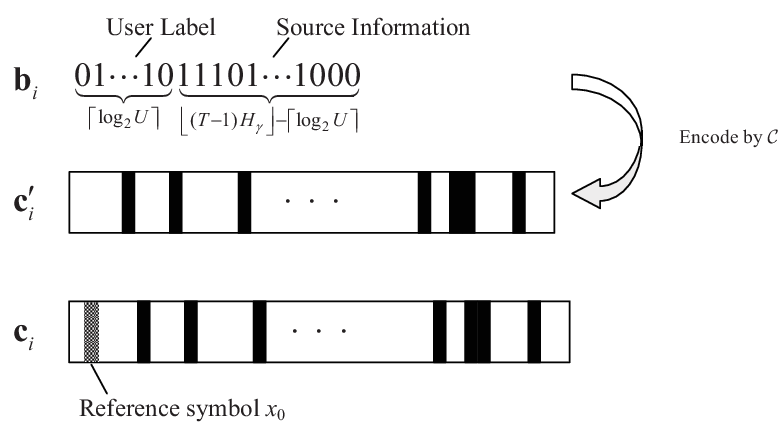}\\
	\caption{Sparse packet generation process of active user $i \in \mathcal{N}$. For the modulated signals $\mathbf{c}'_{i}$ and $\mathbf{c}_i$, the zero and non-zero symbols are represented by white and black rectangles, respectively.}\label{fig_packet_generation}
\end{figure}

For each $i \in \mathcal{N}$, user $i$ first generates a binary data stream $\mathbf{b}_i$ consisting of $\lfloor (T-1)H_\gamma \rfloor$ bits, where
\begin{equation}
\label{eq_symbol_entropy}
H_{\gamma} = - (1-\gamma)\log_2(1-\gamma) - \gamma \log_2\left(\frac{\gamma}{|\mathcal{X}|}\right)
\end{equation}
is the number of information bits carried by each symbol with distribution \eqref{eq_symbol_distribution_TS}. The first $\lceil\log_2 U \rceil$ bits of $\mathbf{b}_i$ represent user $i$'s identity, and the remaining bits are the source information that user $i$ intends to transmit to the BS. Then, the binary stream $\mathbf{b}_i$ is mapped to a modulated signal vector $\mathbf{c}'_i\in \mathbb{C}^{T-1}$ according to a codebook $\mathcal{C} \subset \mathbb{C}^{T-1}$. The codebook $\mathcal{C}$ consists of $2^{\lfloor (T-1)H_\gamma \rfloor}$ distinct codewords with symbols i.i.d. drawn from \eqref{eq_symbol_distribution_TS}, and is assumed to be known at the BS and all users. To produce the final data packet $\mathbf{c}_i$, a reference symbol $x_0$, which is a fixed constellation point in $\mathcal{X}$, is randomly inserted before the first non-zero symbol of $\mathbf{c}_i'$. This is to eliminate the phase ambiguity in packet recovery, as will be detailed in Section \ref{subsection_packet_recovery_NT}. The whole sparse signal generation process is illustrated in Fig. \ref{fig_packet_generation}. Note that the data packet $\mathbf{c}_i$ is scaled appropriately to satisfy the power constraint \eqref{eq_power_TS} before it is transmitted to the BS.

Let $N \triangleq |\mathcal{N}|$ be the number of active users in the considered time slot, assumed to be known at the BS. Further, denote by $i_n$ the $n$-th element of $\mathcal{N}$, $n = 1,2,\cdots,N$. Then, we rewrite the received signal \eqref{eq_received_signal_TS} as
\begin{equation}
\label{eq_problem_TS}
\mathbf{Y} = \mathbf{HX}+\mathbf{W}
\end{equation}
where
\begin{equation}
\mathbf{H} = \left[\mathbf{h}_{i_1},\mathbf{h}_{i_2},\cdots,\mathbf{h}_{i_N}\right] \in \mathbb{C}^{M \times N}
\end{equation}
and
\begin{equation}
\mathbf{X} = \left[\mathbf{c}_{i_1},\mathbf{c}_{i_2},\cdots,\mathbf{c}_{i_N}\right]^T \in \mathbb{C}^{N \times T}
\end{equation}
are the effective channel and signal matrices, respectively. With sparse packet generation described above, $\mathbf{X}$ is a sparse matrix with sparsity level $\gamma$. Moreover, since the symbols in the user packets are generated i.i.d. from \eqref{eq_symbol_distribution_TS}, the entries in $\mathbf{X}$ are independent with each other. That is, the non-zero entries are independently distributed in $\mathbf{X}$. We refer to this property as \textit{random sparsity}.

In RSL-MUD, the BS jointly estimates $(\mathbf{H},\mathbf{X})$ following the maximum a posteriori probability (MAP) principle. Conditioning on $\mathbf{Y}$, the joint posterior pdf of $(\mathbf{H,X})$ is given by
\begin{align}
\label{eq_posterior_pdf_TS}
p_{\mathsf{H,X}|\mathsf{Y}}(\mathbf{H,X}|\mathbf{Y}) &\propto p_{\mathsf{Y}|\mathsf{H,X}}(\mathbf{Y}|\mathbf{H,X})p_{\mathsf{H}}(\mathbf{H})p_{\mathsf{X}}(\mathbf{X})\nonumber\\
&= p_{\mathsf{W}}(\mathbf{Y}-\mathbf{HX})p_{\mathsf{H}}(\mathbf{H})p_{\mathsf{X}}(\mathbf{X})\nonumber\\
&\propto \exp\!\left(\!-\frac{||\mathbf{Y-HX}||_F^2}{\sigma^2}\!\right) p_{\mathsf{H}}(\mathbf{H})p_{\mathsf{X}}(\mathbf{X}).
\end{align}
The MAP estimate of $(\mathbf{H,X})$ is then given by
\begin{equation}
\label{eq_MAP_TS}
\left(\mathbf{\hat{H},\hat{X}}\right)=\mathop{\arg\max}_{(\mathbf{H,X})}\exp\!\left(\!-\frac{||\mathbf{Y\!-\!HX}||_F^2}{\sigma^2}\!\right) p_{\mathsf{H}}(\mathbf{H})p_{\mathsf{X}}(\mathbf{X}).
\end{equation}

We note that the MAP estimation in \eqref{eq_MAP_TS} is, in general, highly complex. However, given that $\mathbf{X}$ is randomly sparse, the joint estimation problem under concern can be seen as a dictionary learning problem in the context of CS. That is, the BS attempts to learn a dictionary matrix $\mathbf{H}$ and the corresponding sparse representation $\mathbf{X}$, based on a noisy observation of the matrix product $\mathbf{HX}$, i.e., $\mathbf{Y}$. In RSL-MUD, we apply a state-of-the-art dictionary learning algorithm, termed BiG-AMP, on the joint estimation problem.

\subsection{Joint Estimation via BiG-AMP}
\label{subsection_blinear}
\begin{figure}
	\center
	\includegraphics[width=7cm]{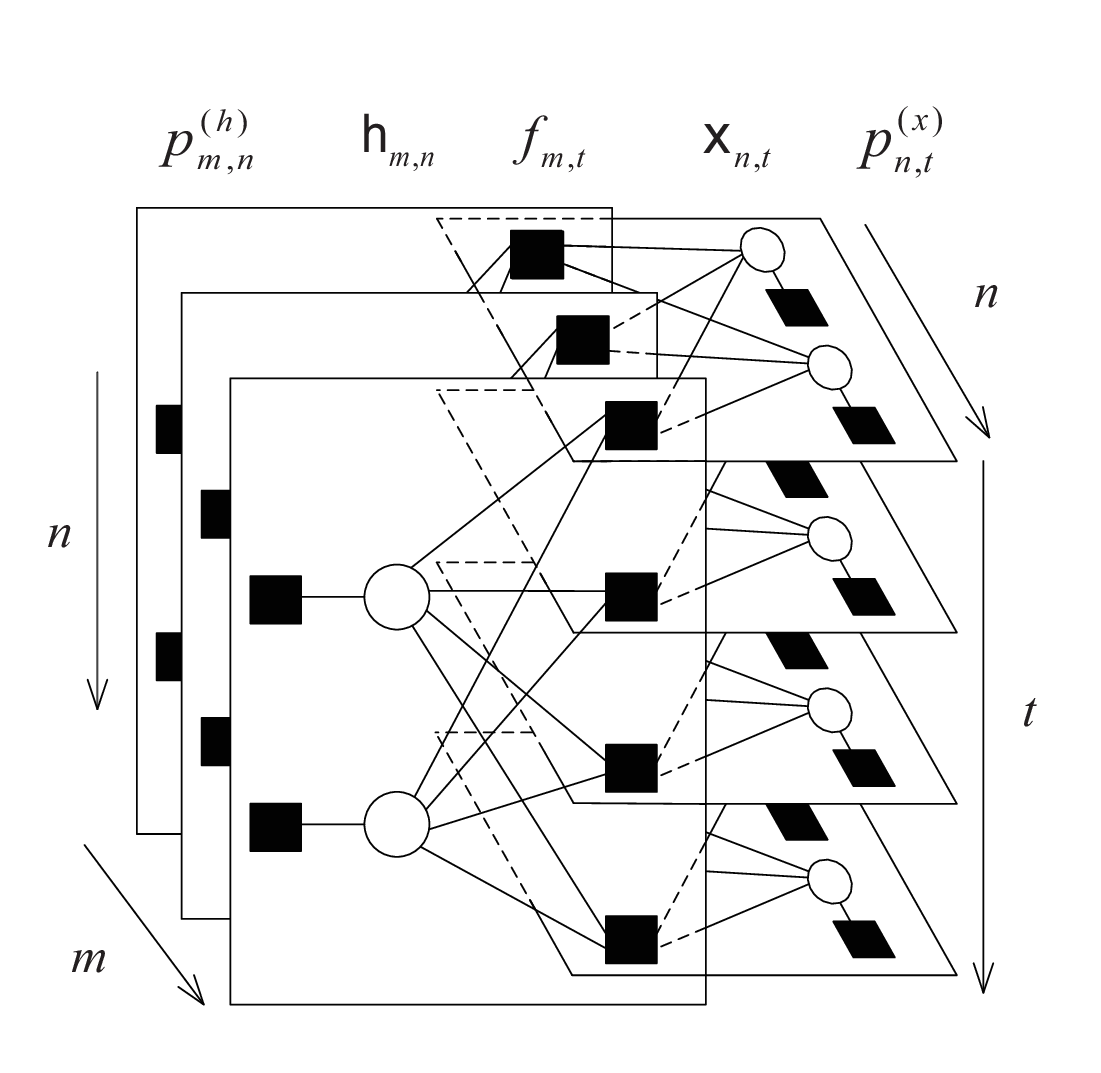}\\
	\caption{The factor graph for the BiG-AMP algorithm with toy-problem parameters $M = 3$, $N = 2$, and $T = 4$. Variable and factor nodes are represented by hollow circles and solid squares, respectively.}\label{fig_factor_graph_TS}
\end{figure}

In this subsection, we describe the BiG-AMP algorithm for the joint estimation of $(\mathbf{H},\mathbf{X})$. First, we rewrite the constraint \eqref{eq_problem_TS} as
\begin{equation}
\label{eq_bilinear_TS}
y_{m,t} = \sum^N_{n=1}h_{m,n}x_{n,t} + w_{m,t},\quad \forall m,t.
\end{equation}
The constraints in \eqref{eq_bilinear_TS} are bilinear with respect to $\{h_{m,n},x_{n,t}\}$, and can be modeled in a factor graph shown in Fig. \ref{fig_factor_graph_TS}. The factor graph consists of two types of nodes, i.e., variable nodes and factor nodes. The variable nodes include $\{\mathsf{h}_{m,n}\}$ and $\{\mathsf{x}_{n,t}\}$. The factor nodes include $\{f_{m,t}\}$ for the constraints in \eqref{eq_bilinear_TS}, $\{p^{(x)}_{n,t}\}$ for the prior distributions $\{p_{\mathsf{x}_{n,t}}(x_{n,t})\}$, and $\{p^{(h)}_{m,n}\}$ for the prior distributions $\{p_{\mathsf{h}_{m,n}}(h_{m,n})\}$.

The inference problem in Fig. \ref{fig_factor_graph_TS} can be solved by the BiG-AMP algorithm \cite{ref_BiGAMP}. The BiG-AMP algorithm approximates the marginal distributions of $\mathsf{h}_{m,n}$ and $\mathsf{x}_{n,t}$ respectively as
\begin{align}
\label{eq_posterior_h}
&p_{\mathsf{h}_{m,n}|\mathsf{Y}}\left(h_{m,n}|\mathbf{Y}\right) = \nonumber\\ &\quad \quad\quad \frac{p_{\mathsf{h}_{m,n}}(h_{m,n})\mathcal{CN}(h_{m,n};\hat{q}_{m,n},\nu^{(q)}_{m,n})}{ \int    p_{\mathsf{h}_{m,n}}\!(h_{m,n})\mathcal{CN}(h_{m,n};\hat{q}_{m,n},\nu^{(q)}_{m,n})dh_{m,n}}
\end{align}
and
\begin{align}
\label{eq_posterior_x}
p_{\mathsf{x}_{n,t}|\mathsf{Y}}\left(x_{n,t}|\mathbf{Y}\right) = \frac{p_{\mathsf{x}_{n,t}}(x_{n,t})\mathcal{CN}(x_{n,t};\hat{r}_{n,t},\nu^{(r)}_{n,t})}{ \int    p_{\mathsf{x}_{n,t}}(x_{n,t})\mathcal{CN}(x_{n,t};\hat{r}_{n,t},\nu^{(r)}_{n,t})dx_{n,t}}
\end{align}
where the parameters $\hat{q}_{m,n},\nu^{(q)}_{m,n},\hat{r}_{n,t},\nu^{(r)}_{n,t}$ are iteratively updated. For completeness, the BiG-AMP algorithm is summarized in Table \ref{tab_BiGAMP}. A brief explanation of the algorithm is provided in the following.

In the initialization, we compute the means and variances of $\{\mathsf{h}_{m,n},\mathsf{x}_{n,t}\}$ according to the prior distributions $\{p_{\mathsf{h}_{m,n}}(h_{m,n})\}$ and $\{p_{\mathsf{x}_{n,t}}(x_{n,t})\}$. In (A1)-(A2) of Table I, the messages from the variable nodes $\{\mathsf{h}_{m,n},\mathsf{x}_{n,t}\}$ to the factor nodes $\{f_{m,t}\}$ are accumulated to obtain an estimate of $\mathbf{HX}$ with means $\{\bar{w}_{m,t}(l)\}$ and variances $\{\bar{\nu}^w_{m,t}(l)\}$. In (A3)-(A4), in order to calculate the output messages of factor nodes $\{f_{m,t}\}$, the Onsager correction \cite{ref_Onsager} is applied to generate the adjusted means $\{\hat{w}_{m,t}(l)\}$ and variances $\{\nu^w_{m,t}(l)\}$. In (A5)-(A12), the Onsager corrected means and variances are used to generate messages that passed from the factor nodes $\{f_{m,t}\}$ to the variable nodes $\{\mathsf{h}_{m,n},\mathsf{x}_{n,t}\}$. Specifically, for each $\mathsf{h}_{m,n}$, a mean $\hat{q}_{m,n}(l)$ and the corresponding variance $\nu^q_{m,n}(l)$ are computed, while for each $\mathsf{x}_{n,t}$, a mean $\hat{r}_{n,t}(l)$ and the corresponding variance $\nu^r_{n,t}(l)$ are computed. Then, in (A13)-(A14), each pair of $\hat{q}_{m,n}(l)$ and $\nu^q_{m,n}(l)$ are merged with the prior distribution $p_{\mathsf{h}_{m,n}}(h_{m,n})$ to produce the posterior mean $\hat{h}_{m,n}$ and variance $\nu^h_{m,n}$, where the expectation is taken with respect to the distribution \eqref{eq_posterior_h}. A similar process is performed on each $\mathsf{x}_{n,t}$ in (A15)-(A16), where the expectation is taken with respect to the distribution \eqref{eq_posterior_x}. Finally, in (A17)-(A18), the BiG-AMP algorithm outputs the estimated distributions $\{\hat{p}_{h_{m,n}}(h_{m,n})\}$ and $\{\hat{p}_{x_{n,t}}(x_{n,t})\}$. Note that though not included in Table \ref{tab_BiGAMP}, adaptive damping is required to ensure the convergence of the BiG-AMP algorithm. We refer the interested readers to \cite{ref_BiGAMP} for details.

For the input of the BiG-AMP algorithm, we set the prior distribution of each $\mathsf{x}_{n,t}$ to be
\begin{equation}
p_{\mathsf{x}_{n,t}}(x_{n,t}) = (1 - \gamma) \delta(x) + \frac{\gamma}{|\mathcal{X}|}\delta(x-x_{n,t}),
\end{equation}
which gives the same distribution as \eqref{eq_symbol_distribution_TS}. Since the BS does not know the identities of the active users, the prior distribution of each $\mathsf{h}_{m,n}$ is set to be $\mathcal{CN}(0,\bar{\beta})$, where $\bar{\beta} = \sum^U_{i=1}\beta_i / U$ is the average path-loss and shadowing component of all users in the system.

\begin{table}
	\caption{BiG-AMP Algorithm}
	\label{tab_BiGAMP}
	\begin{center}
		\begin{tabular}{ |c| }
			\hline
			\includegraphics[scale=0.65]{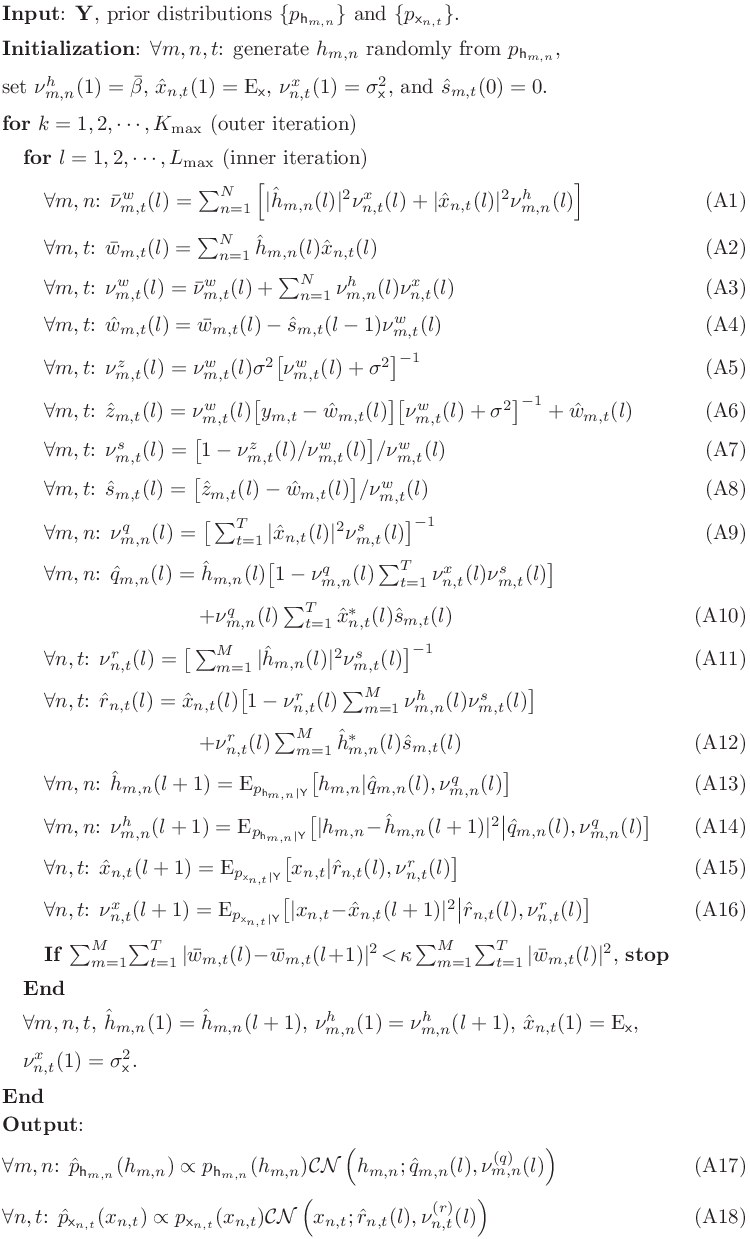}\\
			\hline
		\end{tabular}
	\end{center}
\end{table}

\subsection{Packet Recovery and Data Decoding}
\label{subsection_packet_recovery_NT}
In this subsection, we describe the ambiguity elimination, the packet recovery, and the data decoding procedure based on the output of the BiG-AMP algorithm.

The estimation problem \eqref{eq_MAP_TS} suffers from permutation and phase ambiguities \cite{ref_blind}, which should be handled appropriately before packet recovery. Specifically, the distribution \eqref{eq_posterior_pdf_TS} is invariant to row permutation and certain phase shift of $\mathbf{X}$. For the modulation alphabet $\mathcal{X}$, we define
\begin{equation}
\mathcal{S}_{\mathcal{X}} = \{a : ax\in\mathcal{X}, \forall x \in \mathcal{X}\}.
\end{equation}
That is, for any $x \in \mathcal{X}$ and $a \in \mathcal{S}_\mathcal{X}$, $ax$ is still contained in $\mathcal{X}$. Let $\mathbf{\Sigma}$ be a diagonal matrix with the entries selected from $\mathcal{S}_\mathcal{X}$, and $\mathbf{\Pi}$ be an arbitrary permutation matrix. Then the ambiguities can be seen from the fact that, if $(\hat{\mathbf{H}},\hat{\mathbf{X}})$ is a solution to \eqref{eq_MAP_TS}, $(\mathbf{H},\mathbf{X}) = (\hat{\mathbf{H}}\mathbf{\Pi}^{-1}\mathbf{\Sigma}^{-1},\mathbf{\Sigma}\mathbf{\Pi}\hat{\mathbf{X}})$ is also a valid solution to \eqref{eq_MAP_TS}.

Note that the permutation ambiguity does not affect the packet recovery. The reason is that the user identity information is contained in each packet, and therefore given any row-permutated matrix of $\mathbf{X}$, we can still recover all the user packets and identify their transmitters. The phase ambiguity can be eliminated by exploiting the first non-zero symbol in each packet, which is fixed to be $x_0$. Specifically, we first generate an estimate of each $\mathsf{x}_{n,t}$ as
\begin{equation}
\label{eq_soft_estimate_TS}
\hat{x}^{(s)}_{n,t} = \begin{cases}
\mathrm{E}_{\mathsf{x}_{n,t}}[x_{n,t}], & \left|\mathrm{E}_{\mathsf{x}_{n,t}}[x_{n,t}]\right| \geq \epsilon\\
0, &\text{otherwise}
\end{cases}
\end{equation}
where the expectation is taken with respect to the output distribution $\hat{p}_{\mathsf{x}_{n,t}}(x_{n,t})$ of the BiG-AMP algorithm. Note that in \eqref{eq_soft_estimate_TS} we estimate an entry to be zero if
its expectation is smaller than a threshold $\epsilon$. Then for each $n$, we scale $\{\hat{x}^{(s)}_{n,t}\}^{T}_{t=1}$ such that the first non-zero entry is estimated to be $x_0$ exactly, i.e.
\begin{equation}
\label{eq_soft_estimate2_TS}
\hat{x}^{(c)}_{n,t} = \frac{x_0 \hat{x}^{(s)}_{n,t}}{\hat{x}^{(s)}_{n,\hat{\tau}_n}}, \quad \!\! t=1,2, \cdots, T
\end{equation}
where
\begin{equation}
\hat{\tau}_n = \min \{t : \hat{x}^{(s)}_{n,t} \not= 0\}
\end{equation}
is the estimated index of the first non-zero entry in the $n$-row. In this way, the phase ambiguity is eliminated.

Finally, we perform a hard decision on each $\mathsf{x}_{n,t}$ as
\begin{equation}
\hat{x}^{(h)}_{n,t} = \begin{cases}
\mathop{\arg\min}\limits_{x\in\mathcal{X}}\left|\hat{x}^{(c)}_{n,t} - x\right|^2, & \hat{x}^{(c)}_{n,t}\not = 0\\
0, & \text{otherwise}
\end{cases}
\end{equation}
for $t=1,2, \cdots, T$, $n = 1,2, \cdots, N$. The transmitted packets are estimated to be
\begin{equation}
\hat{\mathbf{c}}_{n} = \left(\hat{x}^{(h)}_{n,1 },\hat{x}^{(h)}_{n,2 },\cdots, \hat{x}^{(h)}_{n,T}\right)^T, \quad \!\! n = 1,2,\cdots,N.
\end{equation}

To decode the user data, we eliminate the first non-zero symbol in each $\hat{\mathbf{c}}_n$, yielding $\hat{\mathbf{c}}'_n$. We then decode $\hat{\mathbf{c}}'_n$ according to codebook $\mathcal{C}$. Specifically, if $\hat{\mathbf{c}}'_n$ is not a codeword in $\mathcal{C}$, decoding failure is declared. Otherwise, we can recover the binary data stream $\mathbf{b}_i$ of each active user. The active users can be identified by inspecting the first $\lceil \log_2 U \rceil$ bits of each $\mathbf{b}_i$, while the remaining bits of $\mathbf{b}_i$ carry the desired information from the transmitting user.

\section{SSL-MUD for Non-Time-Slotted Grant-Free MaDMA}
In this section, we develop the SSL-MUD for the non-time-slotted grant-free MaDMA. In non-time-slotted grant-free MaDMA, since the user packets are not align into time slots, we can not directly factorize the received signal in the form of \eqref{eq_problem_TS}. In SSL-MUD, we introduce a sliding-window-based detection framework. We note that within a relatively long time window, the received signal at the BS can be factorize as the product of an effective channel matrix and an effective signal matrix, where the effective signal matrix naturally exhibits a \textit{structured sparsity}. Exploiting this structured sparsity, we put forth a message passing based algorithm, termed Turbo-BiG-AMP, to recover the packets that completely fall in the considered window. In this way, the BS can recover all the user packets over a continuously moving time window.

\subsection{Sliding-Window Framework}
Different from RSL-MUD, in SSL-MUD we do not require users to transmit sparse packets. The transmitted symbols of active users are simply generated from a modulation constellation $\mathcal{X}$ with $0\not\in\mathcal{X}$. Each packet carries the identity and the source information of the transmitting user. Further, as in RSL-MUD, the first symbol of each packet is set to be a fixed constellation point $x_0 \in \mathcal{X}$, so as to eliminate the phase ambiguity in packet recovery.

Consider an arbitrary observation window $[t_0,t_0+T')$  lasting $T'$ symbol durations with $T' > L$. For each user packet $\mathbf{c}^{(j)}_{i}$, we define
\begin{equation}
\tilde{c}^{(j)}_{i,t} = \begin{cases}
c^{(j)}_i(t-t^{(j)}_i + 1), & t^{(i)}_j \leq t < t^{(i)}_j + L\\
0, & \text{otherwise}
\end{cases}
\end{equation}
and let
\begin{equation}
\label{eq_def_xrow}
\mathbf{x}^{(j)}_i = \left[\tilde{c}^{(j)}_{i,t_0}, \tilde{c}^{(j)}_{i,t_0+1}, \cdots, \tilde{c}^{(j)}_{i,t_0+T'-1}\right]^T \in \mathbb{C}^{T'\times 1}
\end{equation}
be the portion of packet $\mathbf{c}^{(j)}_i$ fallen into the observation window. Then the received signal matrix at the BS during the observation window is
\begin{align}
\label{eq_received_signal}
\mathbf{Y}_{t_0} &=\left[\mathbf{y}_{t_0},\cdots,\mathbf{y}_{t_0+T'-1}\right] \nonumber \\
&=  \sum_{(i,j):i\in \mathcal{I}_U\atop j>0}\mathbf{h}_i^{(j)}\left(\mathbf{x}_{i}^{(j)}\right)^T\!+\!\mathbf{W}_{t_0}  \in \mathbb{C}^{M \times T'}
\end{align}
where
\begin{equation}
\mathbf{W}_{t_0}=\left[\mathbf{w}_{t_0},\mathbf{w}_{t_0+1},\cdots,\mathbf{w}_{t_0+T'-1}\right] \in \mathbb{C}^{M \times T'}.
\end{equation}

\begin{figure}
	\center
	\includegraphics[width=8cm]{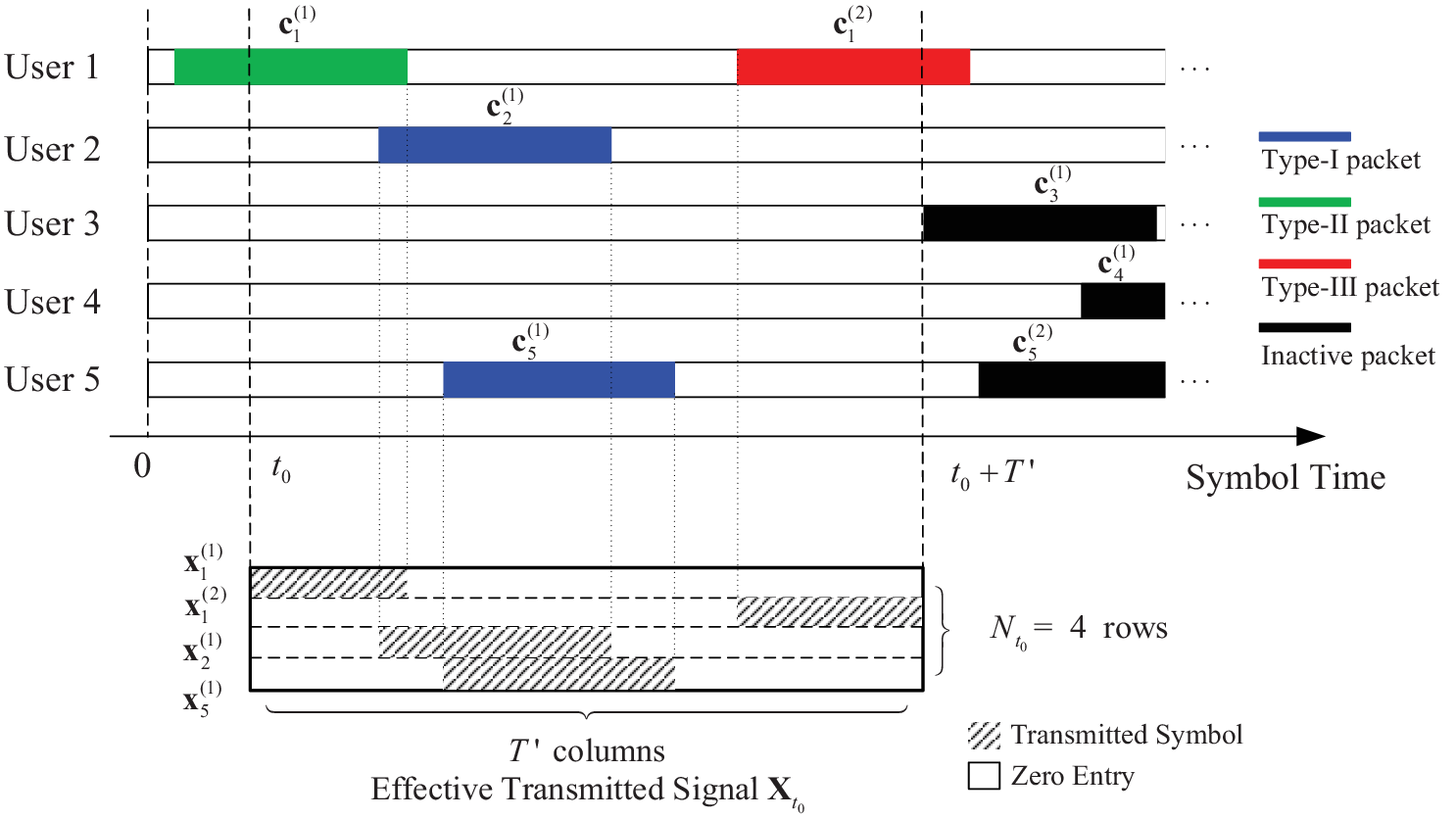}\\
	\caption{An example of packets transmitted in the non-time-slotted grant-free MaDMA. Although 5 users transmit 7 packets in total, only 4 packets, i.e., $\mathbf{c}^{(1)}_1,\mathbf{c}^{(2)}_1,\mathbf{c}^{(1)}_2,\mathbf{c}^{(1)}_5$, are active in the observation window $[t_0,t_0+T')$, and the effective transmitted signal $\mathbf{X}_{t_0}$ has 4 rows. Further, user 1 has 2 active packets, corresponding to 2 different rows in $\mathbf{X}_{t_0}$.}\label{fig_packet_type}
\end{figure}

As illustrated in Fig. \ref{fig_packet_type}, a particular packet $\mathbf{c}^{(j)}_i$ may be out of the window, within the window, or across the boundary of the window, resulting in different patterns of $\mathbf{x}^{(j)}_i$. Further, due to the random transmission of user packets, only a small number of ``active" packets appear in the observation window $[t_0,t_0+T')$. Most of the packets are out of the window with the corresponding $\mathbf{x}^{(j)}_i = \mathbf{0}$. Therefore, for the observation window, it suffices to consider the packets with $\mathbf{x}^{(j)}_i \not= \mathbf{0}$. Define a set of index pairs corresponding to all the active packets by
\begin{equation}
\mathcal{A}_{t_0} \triangleq \left\{(i,j): t_0-L+1 \leq t^{(j)}_i < t_0 + T'\right\}.
\end{equation}
For any $(i,j)\in \mathcal{A}_{t_0}$, user $i$ transmits either the whole packet $\mathbf{c}^{(j)}_i$ or a portion of $\mathbf{c}^{(j)}_i$ during the observation window. Denote by $(i_a,j_a)$ the $a$-th element of $\mathcal{A}_{t_0}$ for $a=1,\cdots,N_{t_0}$, where $N_{t_0} \triangleq |\mathcal{A}_{t_0}|$ is the number of active packets in the observation window, assumed to be known at the BS. Then we rewrite \eqref{eq_received_signal} as
\begin{equation}
\label{eq_channel_model}
\mathbf{Y}_{t_0} = \mathbf{H}_{t_0}\mathbf{X}_{t_0} + \mathbf{W}_{t_0}
\end{equation}
where
\begin{equation}
\mathbf{H}_{t_0} = \left[\mathbf{h}^{(j_1)}_{i_1},\cdots,\mathbf{h}^{(j_{N_{t_0}})}_{i_{N_{t_0}}}\right] \in \mathbb{C}^{M \times N_{t_0}}
\end{equation}
and
\begin{equation}
\mathbf{X}_{t_0} = \left[\left(\mathbf{x}^{(j_1)}_{i_1}\right), \cdots, \left(\mathbf{x}^{(j_{N_{t_0}})}_{i_{N_{t_0}}}\right)\right]^T \in \mathbb{C}^{N_{t_0} \times T'}
\end{equation}
are the effective channel and transmitted signal during the observation window, respectively. The columns of $\mathbf{H}_{t_0}$ are the channel vectors corresponding to different active packets, and hence are independent of each other. The rows of $\mathbf{X}_{t_0}$ contain all the active packets that appear in the observation window. Based on whether a packet is within the window or across the boundary of the window, we categorize the packets in $\mathcal{A}_{t_0}$ into three types:
\begin{itemize}
	\item
	\textbf{Type I}: packet $\mathbf{c}^{(j)}_i$ is within the observation window, i.e., $t_0 \leq t^{(j)}_i \leq t_0 + T' - L$;
	\item
	\textbf{Type II}: packet $\mathbf{c}^{(j)}_i$ crosses the left boundary of the observation window, i.e., $ t_0- L + 1 \leq t^{(j)}_i < t_0$;
	\item
	\textbf{Type III}: packet $\mathbf{c}^{(j)}_i$ crosses the right boundary of the observation window, i.e., $ t_0 + T' - L < t^{(j)}_i < t_0 + T'$.
\end{itemize}
An example of $\mathbf{X}_{t_0}$ and the three types of packets are shown in Fig. \ref{fig_packet_type}. Since each row of $\mathbf{X}_{t_0}$ corresponds to one packet and has at most $L$ non-zero entries, the ratio of non-zero entries of $\mathbf{X}_{t_0}$ is upper bounded by $L/T'$. As a result, the effective signal $\mathbf{X}_{t_0}$ is essentially sparse, provided that $T'$ is sufficiently greater than the packet length $L$.

For a particular observation window $[t_0,t_0+T')$, the BS aims to jointly estimate $\mathbf{H}_{t_0}$ and $\mathbf{X}_{t_0}$ based on $\mathbf{Y}_{t_0}$ in \eqref{eq_channel_model}. Once $\mathbf{X}_{t_0}$ is estimated, the BS is able to recover all the type-I packets. Based on this idea, we introduce a sliding-window-detection framework described as follows. The BS generates a sequence of observation windows $\left\{[t_k,t_k+T')\right\}_{k \in \mathbb{Z}_+}$ where
\begin{equation}
t_k = \begin{cases}
0,&k=1 \\
t_{k-1}+\Delta t,&k\geq 2.
\end{cases}
\end{equation}
This sequence can be seen as a sliding window with window size $T'$ and step size $\Delta t$. Note that for $T'>L$, any two consecutive observation windows have an intersection of $T'-\Delta t$ symbol intervals. It can be verified that if $\Delta t < T' - L$, for each packet $\mathbf{c}^{(j)}_i$, there exists at least one $k\in \mathbb{Z}_+$ such that $\mathbf{c}_i^{(j)}$ is a type-I packet of observation window $[t_k,t_k+T')$. In SSL-MUD, the BS successively estimates the effective transmitted signal $\mathbf{X}_{t_k}$ and decodes all the type-I packets for each observation window $[t_k,t_k+T')$. In this way, the BS eventually recovers all the packets transmitted by the users.

\subsection{Joint Estimation at BS}
In SSL-MUD, the BS jointly estimates the effective channel $\mathbf{H}_{t_k}$ and the transmitted signal $\mathbf{X}_{t_k}$ for each observation window $[t_k,t_k+T')$. In the following, we focus on an arbitrary window $[t_k,t_k+T')$ and omit the subscript $t_k$ in the relevant variables to simplify notation.

Consider the problem of estimating both $\mathbf{H}\in \mathbb{C}^{M \times N}$ and $\mathbf{X} \in \mathbb{C}^{N\times T'}$ from the noisy observation
\begin{equation}
\label{eq_problem_NTS}
\mathbf{Y} = \mathbf{HX}+\mathbf{W} \in \mathbb{C}^{M \times T'}
\end{equation}
where $\mathbf{H}$ is the effective channel, $\mathbf{X}$ is the effective transmitted signal, and $\mathbf{W}$ is the AWGN noise matrix. Although \eqref{eq_problem_NTS} has the same form with \eqref{eq_problem_TS}, there is a significant difference between the joint estimation of $(\mathbf{H},\mathbf{X})$ in RSL-MUD and SSL-MUD. In RSL-MUD, the entries in the sparse signal matrix are independent. However, in SSL-MUD, the sparsity in the effective transmitted signal $\mathbf{X}$ is highly \textit{structured}. That is, each row of $\mathbf{X}$ corresponds to only one active packet in the observation window, and hence the non-zero entries of each row are clustered to form a block with length no greater than $L$, as illustrated in Fig. \ref{fig_packet_type}.

\begin{figure}
	\center
	\includegraphics[width=8cm]{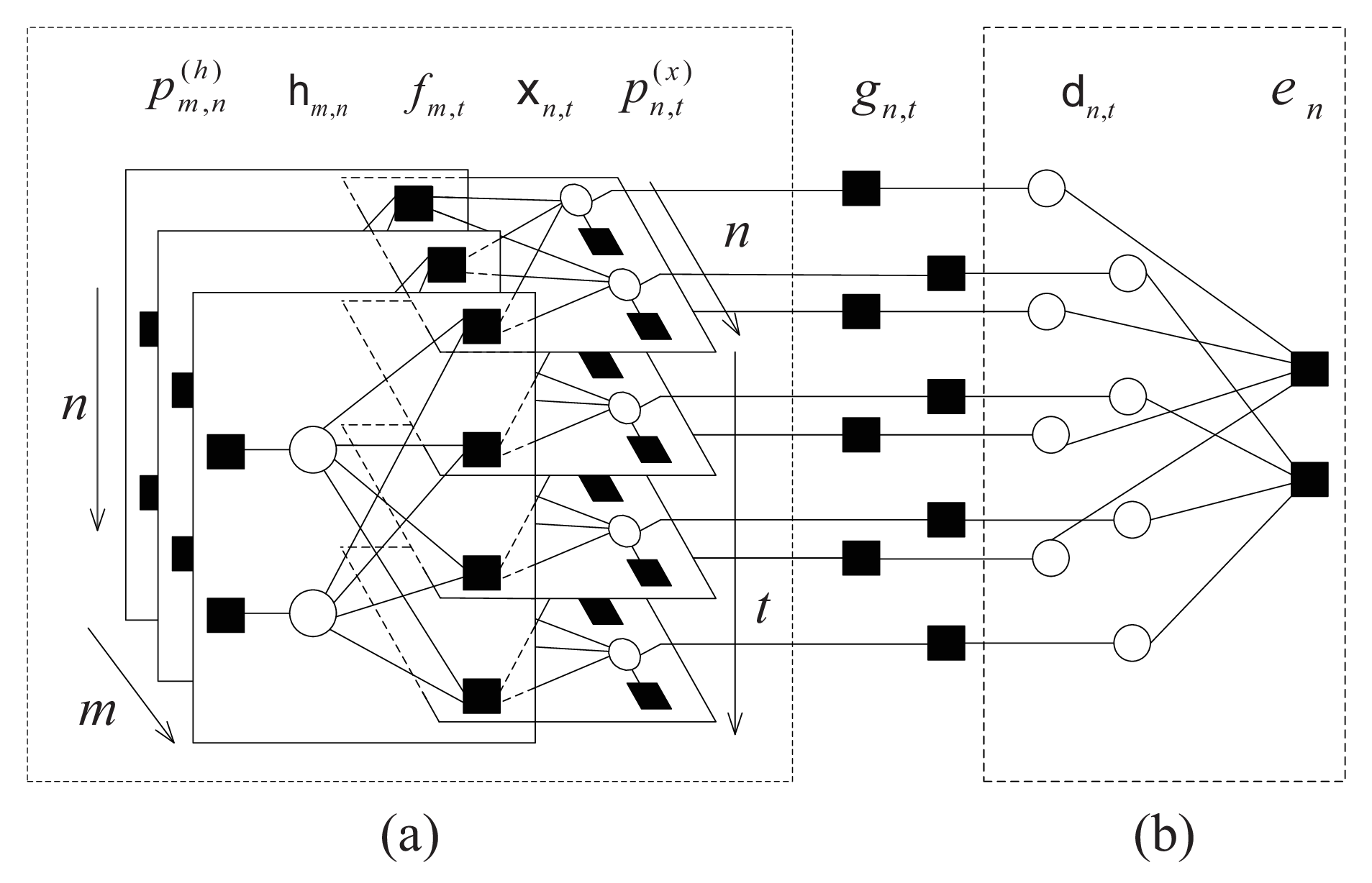}\\
	\caption{The factor graph for the Turbo-BiG-AMP algorithm with toy-problem parameters $M = 3$, $N = 2$, and $T' = 4$. Variable and factor nodes are represented by hollow circles and solid squares, respectively. Subgraph (a) models the bilinear constraint, and subgraph (b) models the structured sparsity.}\label{fig_factor_graph}
\end{figure}

We put forth a joint estimation algorithm, termed Turbo-BiG-AMP, to exploit the structured sparsity in $\mathbf{X}$. The Turbo-BiG-AMP algorithm partitions the factor graph of the joint estimation problem into two subgraphs, as shown in Fig. \ref{fig_factor_graph}. Subgraph (a), referred to as the \textit{bilinear subgraph}, is similar to Fig. \ref{fig_factor_graph_TS} and models the bilinear constraint \eqref{eq_problem_NTS}. Subgraph (b), referred to as the \textit{structured sparsity subgraph}, models the structured sparsity in $\mathbf{X}$. As in RSL-MUD, the inference problem on the bilinear subgraph is solved by the BiG-AMP algorithm. For the structured sparsity subgraph, we refine the estimate on $\mathbf{X}$ by inferring the packet locations, i.e., the locations of non-zero entries in $\mathbf{X}$. The inferences on the two subgraphs are performed iteratively until convergence, and the algorithm outputs a final estimate of $\mathbf{X}$.

\subsection{Structured Sparsity Learning}
\label{subsection_structured_sparsity}
In this subsection, we describe the inference on the packet locations based on the structured sparsity subgraph.

To model the structured sparsity in $\mathbf{X}$, we introduce an auxiliary matrix $\mathbf{D}\in \{0,1\}^{N\times T'}$ to indicate whether the entries in $\mathbf{X}$ are non-zero. Specifically, we let
\begin{equation}
\label{eq_depenency_entry}
p_{\mathsf{x|d}}(x_{n,t}|d_{n,t})\! =\! \begin{cases}
\delta(x_{n,t}),&\!d_{n,t} \!=\! 0\\
p'_{\mathsf{x}}(x_{n,t}), & \!d_{n,t} \!=\! 1
\end{cases}
\end{equation}
where $p'_{\mathsf{x}}(x_{n,t}) = \frac{1}{|\mathcal{X}|} \sum_{x\in \mathcal{X}}\delta(x-x_{n,t})$ is the prior distribution of the transmitted symbol of an active user. That is, $\mathsf{d}_{n,t}=1$ indicates $\mathsf{x}_{n,t}\not=0$ and $\mathsf{d}_{n,t}=0$ indicates $\mathsf{x}_{n,t}=0$. Recall that each row of $\mathbf{X}$ corresponds to an active packet in the observation window. For the $n$-th row of $\mathbf{X}$, let $\Delta t_n$ be the index difference between the first symbol interval of the corresponding packet and the first symbol interval of the observation window. Then, we have
\begin{equation}
-L+1 \leq \Delta t_n \leq T'-1, \quad \!\! \forall 1\leq n \leq N
\end{equation}
Note that an entry of $\mathbf{X}$ is non-zero if it is in an active packet, and is zero otherwise. This implies that for any $n\in\{1,\cdots,N\}$, $\{d_{n,t}\}^{T'}_{t=1}$ are constrained by
\begin{equation}
\label{eq_constraint_deltat}
\mathsf{d}_{n,t} = \begin{cases}
1, & \max\{\Delta t_n + 1, 1\}\leq t \leq \min\{\Delta t_n + L , T'\}\\
0, & \text{otherwise}.
\end{cases}
\end{equation}
The constraint \eqref{eq_constraint_deltat} is modeled as factor nodes $\{e_n\}$ in the structured sparsity subgraph.

The inference on the structured sparsity subgraph is performed by message passing. Let $\Phi^a_b(\cdot)$ represent the message passed from node $a$ to node $b$, and assume that all the messages are scaled to be valid pdfs (for continuous random variables) or probability mass functions (for discrete random variables). First, the variable nodes $\{\mathsf{d}_{n,t}\}^{T'}_{t=1}$ pass their likelihoods to each factor node $e_n$. For each $1\leq t\leq T'$, the distribution of $\Delta t_n$ conditioned on $\{\mathsf{d}_{n,t'}\}^{T'}_{t'=1}\backslash \{\mathsf{d}_{n,t}\}$ is computed as
\begin{align}
\label{eq_distribution_deltat}
p^{(t)}_{\Delta \mathsf{t}_n}(\Delta t_n) \propto &\!\!\!\! \prod^{\max\{\Delta t_n +1,1\}-1 }_{t'=1\atop t' \not= t}\!\!\!\Phi^{\mathsf{d}_{n,t'}}_{e_n}(0)
\times\!\!\!\! \prod^{ \min\{\Delta t_n +L, T'\}}_{t'=\max\{\Delta t_n +1, 1\}\atop t'\not= t}\!\!\!\Phi^{\mathsf{d}_{n,t'}}_{e_n}(1)  \nonumber\\
& \quad \quad \quad\times  \!\!\!\!\!\!\! \prod^{T'}_{t'=\min\{\Delta t_n +L, T'\}+1\atop t'\not= t}\!\!\!\Phi^{\mathsf{d}_{n,t'}}_{e_n}(0).
\end{align}
The message passed from $e_n$ to each $\mathsf{d}_{n,t}$ is given by
\begin{subequations}
	\label{eq_message_etod}
	\begin{align}
	\Phi^{e_n}_{\mathsf{d}_{n,t}}(1)&= \sum^{t-1}_{\Delta t_n = t-L}\!\!\!p^{(t)}_{\Delta \mathsf{t}_n}(\Delta t_n) \\
	\Phi^{e_n}_{\mathsf{d}_{n,t}}(0)&= 1 -  \Phi^{e_n}_{\mathsf{d}_{n,t}}(1).
	\end{align}
\end{subequations}
The distribution \eqref{eq_message_etod} can be seen as the conditional distribution of $\mathsf{d}_{n,t}$ given $\{\mathsf{d}_{n,t'}\}^{T'}_{t' = 1} \backslash \{\mathsf{d}_{n,t}\}$, which is the inference result on the structured sparsity subgraph.

\subsection{Turbo-BiG-AMP Algorithm}
In this subsection, we present the overall Turbo-BiG-AMP algorithm. As shown in Fig. \ref{fig_factor_graph}, we introduce factor nodes $\{g_{n,t}\}$ to model the dependency between $\{\mathsf{x}_{n,t}\}$ and $\{\mathsf{d}_{n,t}\}$, i.e., \eqref{eq_depenency_entry}. Based on the dependency, the messages can be exchanged between subgraphs.

\subsubsection{Messages from bilinear subgraph to structured sparsity subgraph}
Note that the BiG-AMP algorithm is used to perform inference on the bilinear subgraph. As shown in Table \ref{tab_BiGAMP}, the BiG-AMP algorithm outputs an estimated distribution $\hat{p}_{\mathsf{x}_{n,t}}(x_{n,t})$ for each variable node $\mathsf{x}_{n,t}$. The output distribution is directly passed to the factor node $g_{n,t}$, i.e., $\Phi_{g_{n,t}}^{\mathsf{x}_{n,t}}(x_{n,t}) = \hat{p}_{\mathsf{x}_{n,t}}(x_{n,t})$. Then the message passed from $g_{n,t}$ to $\mathsf{d}_{n,t}$ is given by
\begin{equation}
\label{eq_message_gtod}
\Phi^{g_{n,t}}_{\mathsf{d}_{n,t}}(1) = \int_{x\not = 0} \Phi^{\mathsf{x}_{n,t}}_{g_{n,t}}(x)dx,\quad\!\! \Phi^{g_{n,t}}_{\mathsf{d}_{n,t}}(0) = 1- \Phi^{g_{n,t}}_{\mathsf{d}_{n,t}}(1).
\end{equation}
The message in \eqref{eq_message_gtod} is used in the inference on the structured sparsity subgraph as the output message of each $\mathsf{d}_{n,t}$.

\subsubsection{Messages from structured sparsity subgraph to bilinear subgraph}
As shown in Section \ref{subsection_structured_sparsity}, the inference on the structured sparsity subgraph outputs a distribution $\Phi^{e_n}_{\mathsf{d}_{n,t}}(d_{n,t})$ for each $\mathsf{d}_{n,t}$, given in \eqref{eq_message_etod}. Then, each $\mathsf{d}_{n,t}$ passes $\Phi^{e_n}_{\mathsf{d}_{n,t}}(d_{n,t})$ to $g_{n,t}$ directly. Based on the dependency in \eqref{eq_depenency_entry}, we have
\begin{equation}
\Phi^{g_{n,t}}_{\mathsf{x}_{n,t}}(x_{n,t}) = \Phi^{e_n}_{\mathsf{d}_{n,t}}(0)\delta(x_{n,t}) + \Phi^{e_n}_{\mathsf{d}_{n,t}}(1) p'_{\mathsf{x}}(x_{n,t}).
\end{equation}
Then we merge $\Phi^{g_{n,t}}_{\mathsf{x}_{n,t}}(x_{n,t})$ and $p_{\mathsf{x}_{n,t}}(x_{n,t})$ to generate a new prior on $\mathsf{x}_{n,t}$, given by
\begin{equation}
\label{eq_message_xtof}
\Phi^{\mathsf{x}_{n,t}}_{f_{m,t}}(x_{n,t}) \propto \Phi^{g_{n,t}}_{\mathsf{x}_{n,t}}(x_{n,t}) p_{\mathsf{x}_{n,t}}(x_{n,t}).
\end{equation}
The distribution $\Phi^{\mathsf{x}_{n,t}}_{f_{m,t}}(x_{n,t})$ will be used as the new prior of $\mathsf{x}_{n,t}$ in the following BiG-AMP iterations.

The whole procedure of the Turbo-BiG-AMP algorithm is summarized in Table \ref{tab_Turbo_BiGAMP}. In the initialization step, the prior distribution of $\{\mathsf{x}_{n,t}\}$ is given by
\begin{equation}
p_{\mathsf{x}_{n,t}}(x_{n,t}) = (1 - \gamma') \delta(x_{n,t}) + \frac{\gamma'}{|\mathcal{X}|}\sum_{x\in \mathcal{X}}\delta(x_{n,t}-x)
\end{equation}
where
\begin{equation}
\gamma' = \frac{L}{T'+L-1}
\end{equation}
is the estimated sparsity level of $\mathbf{X}$. As in RSL-MUD, the prior distribution $p_{\mathsf{h}_{m,n}}(h_{m,n})$ is set to be $\mathcal{CN}(0,\bar{\beta})$. Further, note that the BiG-AMP algorithm is based on loopy belief propagation and requires multiple iterations to converge, while the structured sparsity subgraph is loop-free and the inference will converge in one iteration. Therefore, we schedule the message passing as follows: the inference on the structured sparsity subgraph is performed once after every $L_{\max}$ inner iterations of the BiG-AMP algorithm.

\begin{table}
	\caption{Turbo-BiG-AMP Algorithm}
	\label{tab_Turbo_BiGAMP}
	\begin{center}
		\begin{tabular}{ |c| }
			\hline
			\includegraphics[scale=0.65]{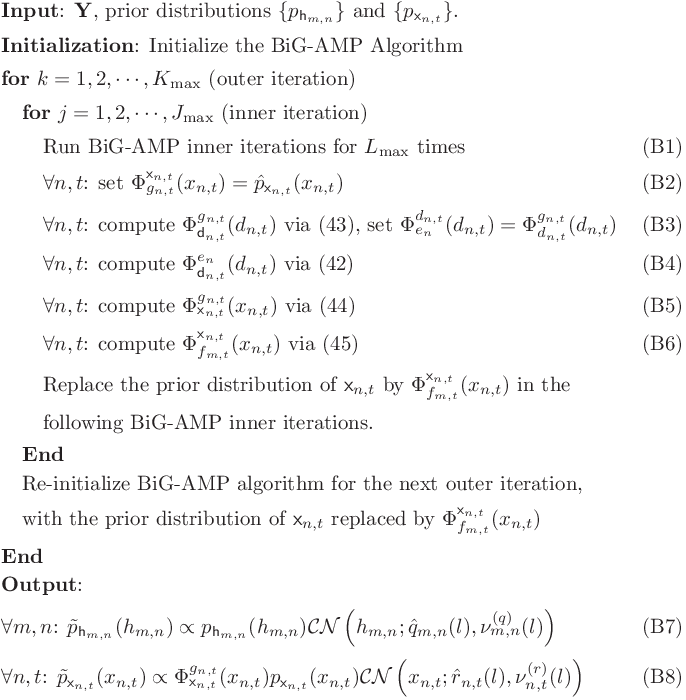}\\
			\hline
		\end{tabular}
	\end{center}
\end{table}

\subsection{Packet Recovery}
\label{subsection_ambiguity}
Based on the output of Turbo-BiG-AMP algorithm, the BS can recover all the type-I packets in the observation window.

\subsubsection{Packet Positioning}
We first determine the locations of non-zero entries in $\mathbf{X}$ for each type-I packet. If the $n$-th row of $\mathbf{X}$ corresponds to a type-I packet, it has exactly $L$ consecutive non-zero entries. In other words, $\Delta t_n$ defined in Section \ref{subsection_structured_sparsity} satisfies
\begin{equation}
0 \leq \Delta t_n \leq T'-L, \quad \!\! \forall 1 \leq n \leq N.
\end{equation}
To estimate $\Delta t_n$, we first compute the distribution of $\mathsf{d}_{n,t}$ as
\begin{subequations}
	\begin{equation}
	\tilde{p}_{\mathsf{d}_{n,t}}(1) = \int_{x\not = 0} \tilde{p}_{\mathsf{x}_{n,t}}(x_{n,t}) dx
	\end{equation}
	\begin{equation}
	\tilde{p}_{\mathsf{d}_{n,t}}(0) = 1 -  \tilde{p}_{\mathsf{d}_{n,t}}(1)
	\end{equation}
\end{subequations}
where $\tilde{p}_{\mathsf{x}_{n,t}}(x_{n,t})$ is the output of the Turbo-BiG-AMP algorithm. Subsequently, we compute the distribution of $\Delta t_n$ conditioned on $\{\mathsf{d}_{n,t}\}^{T'}_{t=1}$ as
\begin{align}
\label{eq_distribution_deltat2}
\tilde{p}_{\Delta \mathsf{t}_n}(\Delta t_n) \propto &\!\!\!\! \prod^{\max\{\Delta t_n +1,1\}-1 }_{t=1}\!\!\!\tilde{p}_{\mathsf{d}_{n,t}}(0) \times \!\!\!\!
\prod^{ \min\{\Delta t_n +L, T'\}}_{t=\max\{\Delta t_n +1, 1\}}\!\!\!\tilde{p}_{\mathsf{d}_{n,t}}(1) \nonumber\\
&\quad \quad \quad\quad \times  \!\!\!\!\!\!\!\prod^{T'}_{t=\min\{\Delta t_n +L, T'\}+1}\!\!\!\tilde{p}_{\mathsf{d}_{n,t}}(0).
\end{align}
Then the estimated $\Delta t_n$ is given by
\begin{equation}
\Delta \hat{t}_n = \mathop{\arg\max}_{-L+1\leq \Delta t_n \leq T'-1} \tilde{p}_{\Delta \mathsf{t}_n}(\Delta t_n).
\end{equation}
The indexes of the rows corresponding to type-I packets are
\begin{equation}
\hat{\mathcal{N}} = \{n:1\leq n\leq N, \quad  0\leq \Delta \hat{t}_n \leq T'-L\}.
\end{equation}

\subsubsection{Phase Ambiguity Elimination}
As in RSL-MUD, the estimation of $\mathbf{X}$ suffers from permutation and phase ambiguities. The permutation ambiguity does not affect the packet recovery, because given any row-permutated matrix of $\mathbf{X}$, we can still obtain all the type-I packets by inspecting the non-zero entries. The phase ambiguity should be eliminated in the packet recovery procedure.

Recall that for each packet, the first symbol is fixed to be $x_0$. For each $n\in\hat{\mathcal{N}}$, we generate a soft estimate of $\mathsf{x}_{n,t}$ as
\begin{equation}
\label{eq_soft_estimate}
\hat{x}^{(s)}_{n,t}\! =\! \frac{x_0 \mathrm{E}_{\mathsf{x}_{n,t}}[x_{n,t}]}{\mathrm{E}_{\mathsf{x}_{n,\Delta \hat{t}_n+1}}[x_{n,\Delta \hat{t}_n + 1}]}, \quad\!\! t \!=\!\Delta \hat{t}_n \!+\! 1, \cdots, \Delta \hat{t}_n \!+\! L.
\end{equation}
where the expectations are taken with respect to the output distribution $\tilde{p}_{\mathsf{x}_{n,t}}(x_{n,t})$ of the Turbo-BiG-AMP algorithm. In \eqref{eq_soft_estimate}, the first symbol of each packet is estimated to be $x_0$, i.e., $\hat{x}^{(s)}_{n,\Delta \hat{t}_n + 1} = x_0$, and therefore the phase ambiguity is eliminated.

Finally, we perform hard decision on each $\mathsf{x}_{n,t}$ as
\begin{equation}
\hat{x}^{(h)}_{n,t}\! = \!
\mathop{\arg\min}\limits_{x\in\mathcal{X}} \big|\hat{x}^{(s)}_{n,t} - x\big|^2, \quad\!\! t =\Delta \hat{t}_n + 1, \cdots, \Delta \hat{t}_n\! +\! L.
\end{equation}
The type-I packets are then decided to be
\begin{equation}
\hat{\mathbf{c}}_{n} = \left(\hat{x}^{(h)}_{n,\Delta \hat{t}+1 },\hat{x}^{(h)}_{n,\Delta \hat{t}+2 },\cdots, \hat{x}^{(h)}_{n,\Delta \hat{t}+L }\right)^T, \quad \!\!\! n\in \hat{\mathcal{N}}.
\end{equation}
The identity of the transmitting user can be extracted from each recovered packet.

\section{Simulation Results}
In this section, we present simulation results to evaluate the performance of the RSL-MUD and SSL-MUD schemes. In the
simulations, we set the number of users $U=200$, the bandwidth to be 1 MHz, and 	the power spectrum density of the AWGN noise at the BS to be -169 dBm/Hz. We assume all active users perform auto gain control (AGC) to ensure an equal average receive power at the BS. In other words, all the active users have identical path-loss and shadowing components, i.e., $\beta_1 = \cdots = \beta_U = \beta$, and we set $\beta = -129$dB. The modulation constellation $\mathcal{X}$ is QPSK constellation, i.e., $\mathcal{X} = \left\{\pm \frac{\sqrt{2}}{2} \pm \frac{\sqrt{2}}{2}i \right\}$. All the simulation results presented in this section are obtained by averaging over 100,000 user packets.

\subsection{RSL-MUD}
In this subsection, we examined the performance of the RSL-MUD scheme for the time-slotted grant-free MaDMA system. For the BiG-AMP algorithm used in RSL-MUD, we set $M_{\max} = 20$, $L_{\max} = 200$, and $\kappa = 10^{-4}$. For each time slot, BiG-AMP is run for $10$ random initializations, and each user packet is considered successfully decoded if it is perfectly recovered in any of the $10$ trials.

\begin{figure}
	\center
	\includegraphics[width=9cm]{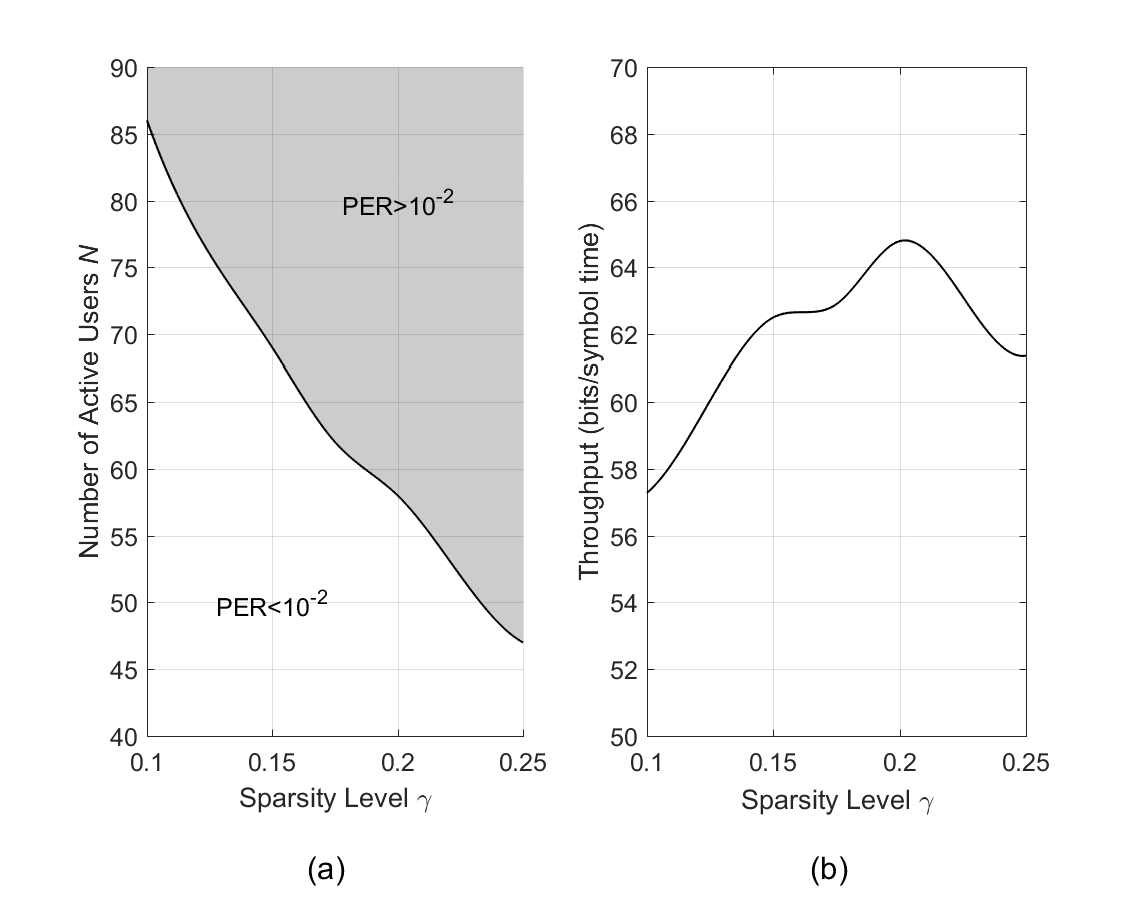}\\
	\caption{For $P = 40$dBm, (a) the phase transition graph of the RSL-MUD scheme with respect to the sparsity level $\gamma$ and the number of active users $N$; (b) the maximum throughput of the RSL-MUD scheme for different sparsity levels. The other system parameters are set as $M = 40$, $T = 256$.}\label{fig_TS_1}
\end{figure}

As stated in Section \ref{section_RSLMUD}, RSL-MUD detects the user packets by exploiting the random sparsity of the signal matrix. It is hence interesting to numerically explore the tradeoff between the system throughput and the sparsity level of the user signals. In particular, we say that RSL-MUD is able to support $N$ active users under sparsity level $\gamma$ if the packet error rate (PER) is no greater than $10^{-2}$. Fig. \ref{fig_TS_1} (a) shows the maximum numbers of active users that can be supported by RSL-MUD under different sparsity levels. We set the number of received antennas $M = 40$, the length of the time slot $T = 256$, and the transmit power $P = 40$dBm. We see that the maximum $N$ is a monotonically decreasing function of the sparsity level $\gamma$. That is, the sparser the user signals, the more the active users that can be supported. However, this does not imply monotonicity of the system throughput against sparsity level $\gamma$, since the increase of sparsity in signaling reduces the information rate of each user. In Fig. \ref{fig_TS_1} (b), we plot the supportable throughput\footnote{The throughput is defined as the effective rate of information bits (excluding the overhead for the reference symbols and the user identity) transmitted to the BS, computed by
	\begin{equation}
	\text{Throughput} = \frac{N \left(\lfloor(T-1)H_{\gamma}\rfloor-\lceil\log_2 U \rceil\right)}{T}.
	\end{equation}
	where $H_{\gamma}$, defined in \eqref{eq_symbol_entropy}, is the information carried by each symbol, and $\lceil\log_2 U \rceil$ is the number of bits used for user identity.}
of RSL-MUD with respect to the sparsity level. It is observed that the supportable throughput is not monotonic with respect to the sparsity level $\gamma$, and achieves its maximum with $\gamma$ around 0.2.

Subsequently, we compare the error performance of RSL-MUD with that of the conventional training-based MUD schemes \cite{ref_CSMUD_noCSI4,ref_CSMUD_noCSI5,ref_CSMUD_noCSI6}. In both schemes, we set $M = 40$ and $T = 256$. For RSL-MUD, we simulate two system setups with $(N,\gamma)$ = $(40,0.25)$ and $(60,0.15)$. For the training-based MUD, we set $N = 40$. Note that in the training-based MUD, each time slot is divided into two phases. In the first phase, $100$ pilot symbols are transmitted by each active user, and the BS jointly estimates the user activity state and the CSI by a state-of-the-art CS approach called generalized approximate message passing (GAMP) \cite{ref_GAMP}. In the second phase, the active users transmit QPSK modulated signals, and the signal detection at the BS is carried out by LMMSE and sphere decoding \cite{ref_sphDec} (with initial searching radius set to 25), respectively.\footnote{In the training-based MUD, user packets are not sparse, and therefore each packet symbol carries more information than that in RSL-MUD. However, for the training-based MUD, a portion of symbols in each packet are used as pilot. In the simulations, system parameters are carefully chosen so that throughputs of RSL-MUD and the training-based MUD are close to each other.} Note that the number of pilot symbols is chosen to ensure a reasonable accuracy of channel estimation, so as to facilitate the signal detection in the second phase. For a fair comparison, in the above settings the throughput of RSL-MUD is no less than that of the training-based MUD.

\begin{figure}
	\center
	\includegraphics[width=8cm]{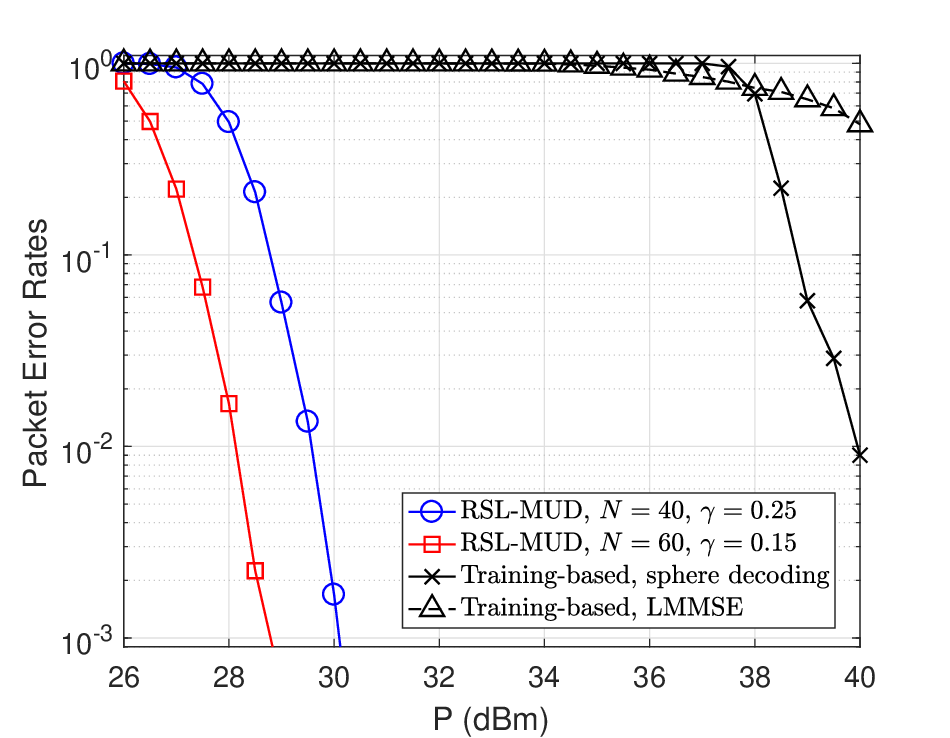}\\
	\caption{PER comparison between RSL-MUD and the training-based MUD. For RSL-MUD, we set $M=40$, $T = 256$, $(N,\gamma)$ = $(40,0.25)$ and $(60,0.15)$. For the training-based MUD, we set $N=40$. In the training-based MUD, the first $100$ symbols of each user packet are used as pilot in the first phase, while in the second phase user signals are decoded by sphere decoding and LMMSE, respectively. For RSL-MUD with $(N,\gamma)=(40,0.25)$ and $(60,0.15)$, the throughputs are 51.05 and 52.59 bits per symbol, respectively. For the training-based MUD, the throughput is $48.75$ bits per symbol.}\label{fig_TS_2}
\end{figure}

Fig. \ref{fig_TS_2} shows the PER results of RSL-MUD and the training-based MUD with respect to the transmit power. We see that for $(N,\gamma) = (40,0.25)$, RSL-MUD achieves a PER of $10^{-2}$ with the transmit power around $29$dBm, and outperforms the training-based MUD with sphere decoding by 10dBm. We also see that the training-based MUD with LMMSE fails to achieve PER lower than $10^{-2}$ in the considered range of transmit power. Further, for $(N,\gamma) = (60,0.15)$, RSL-MUD has a PER gain of 1.5dB compared with $(N,\gamma) = (40,0.25)$. This implies that by appropriately choosing $N$ and $\gamma$, RSL-MUD can achieve a better PER performance without degrading the throughput.

\begin{figure}
	\center
	\includegraphics[width=8cm]{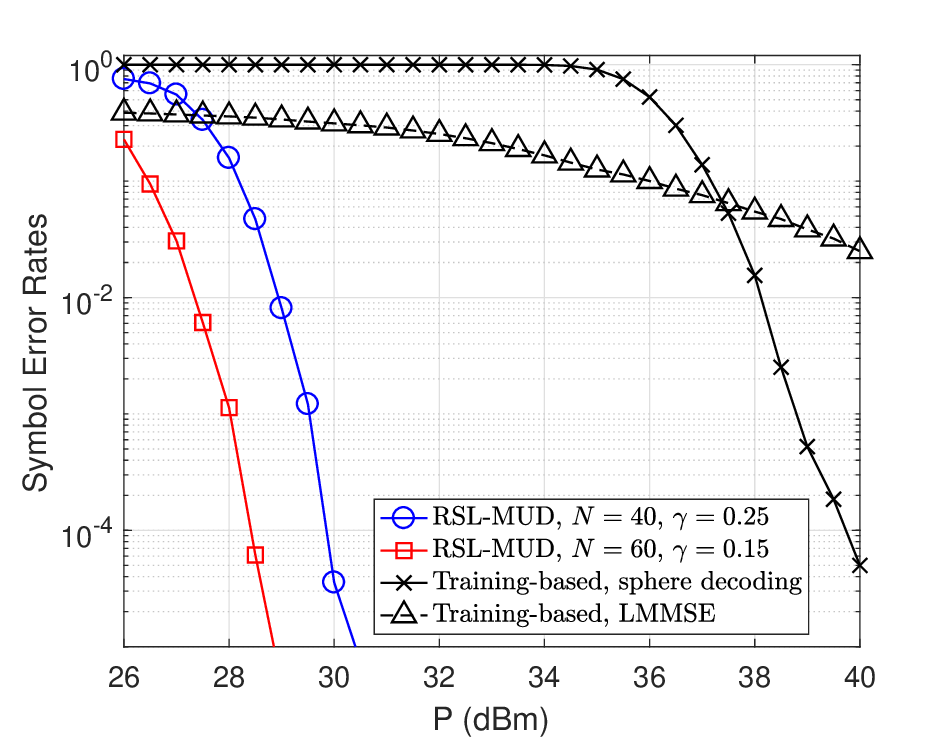}\\
	\caption{SER comparison between RSL-MUD and the training-based MUD.}\label{fig_TS_3}
\end{figure}

\begin{figure}
	\center
	\includegraphics[width=8cm]{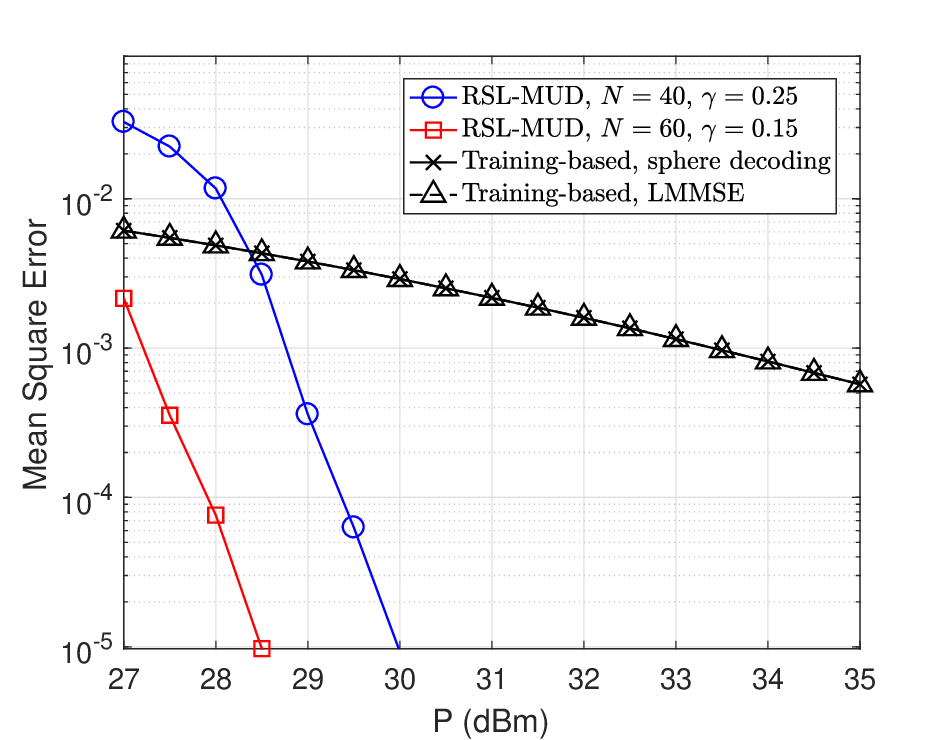}\\
	\caption{MSE of the estimated channels for RSL-MUD and the training-based MUD.}\label{fig_TS_4}
\end{figure}

To further demonstrate the superiority of RSL-MUD, Fig. \ref{fig_TS_3} and Fig. \ref{fig_TS_4} compare the symbol error rates (SER) and the channel estimation accuracy of RSL-MUD and the training-based MUD. For RSL-MUD, if a packet is perfectly recovered, the SER of that packet is set to zero. Otherwise, we record the average SER of 10 trials of the BiG-AMP algorithm. To evaluate the channel estimation accuracy, we define the mean square error (MSE) for an estimated channel $\hat{\mathbf{H}}$ as
\begin{equation}
\text{MSE} = \frac{\min_{\mathbf{Q}}||\mathbf{H}-\hat{\mathbf{H}}\mathbf{Q}||^2_F}{MN}
\end{equation}
where $\mathbf{Q}$ is a permutation matrix to eliminate the column ambiguity of $\hat{\mathbf{H}}$ \cite{ref_BiGAMP}. For RSL-MUD, we use the smallest MSE of 10 trials of the BiG-AMP algorithm. Fig. \ref{fig_TS_3} shows that at the SER level of $10^{-4}$, RSL-MUD with $(N,\gamma) = (40,0.25)$ and $(60,0.15)$ outperforms the training-based MUD by 10dB and 11.5dB, respectively. From Fig. \ref{fig_TS_4}, we see that the channel MSE of RSL-MUD drops significantly faster than that of the training-based MUD. For $(N,\gamma) = (40,0.25)$ and $(60,0.15)$, RSL-MUD achieves a channel MSE of $10^{-3}$ with the transmit power around $27.2$dBm and $28.8$dBm, respectively, while the training-based MUD reaches the same MSE level with $P = 33$dBm.

\subsection{SSL-MUD}
In this subsection, we present simulation results of the SSL-MUD scheme for the non-time-slotted grant-free MaDMA system. We set the length of user packet $L=64$. The size and the step size of the sliding window are given by $T' = 256$ and $\Delta t = 64$, respectively. The data packets of each user are generated according to a Poisson point process with rate $\lambda$. For each user, the guard period between two consecutive packets is set to 64 symbol intervals. Note that a packet is considered successfully decoded if it is perfectly recovered (as a type-I packet) in any observation window. For each observation window, the Turbo-BiGAMP-AMP algorithm is run for only one trial. Parameters of Turbo-BiG-AMP are set as $K_{\max} = 20$, $L_{\max} = 100$, $J_{\max} = 2$, and $\kappa = 10^{-4}$.

\begin{figure}
	\center
	\includegraphics[width=8cm]{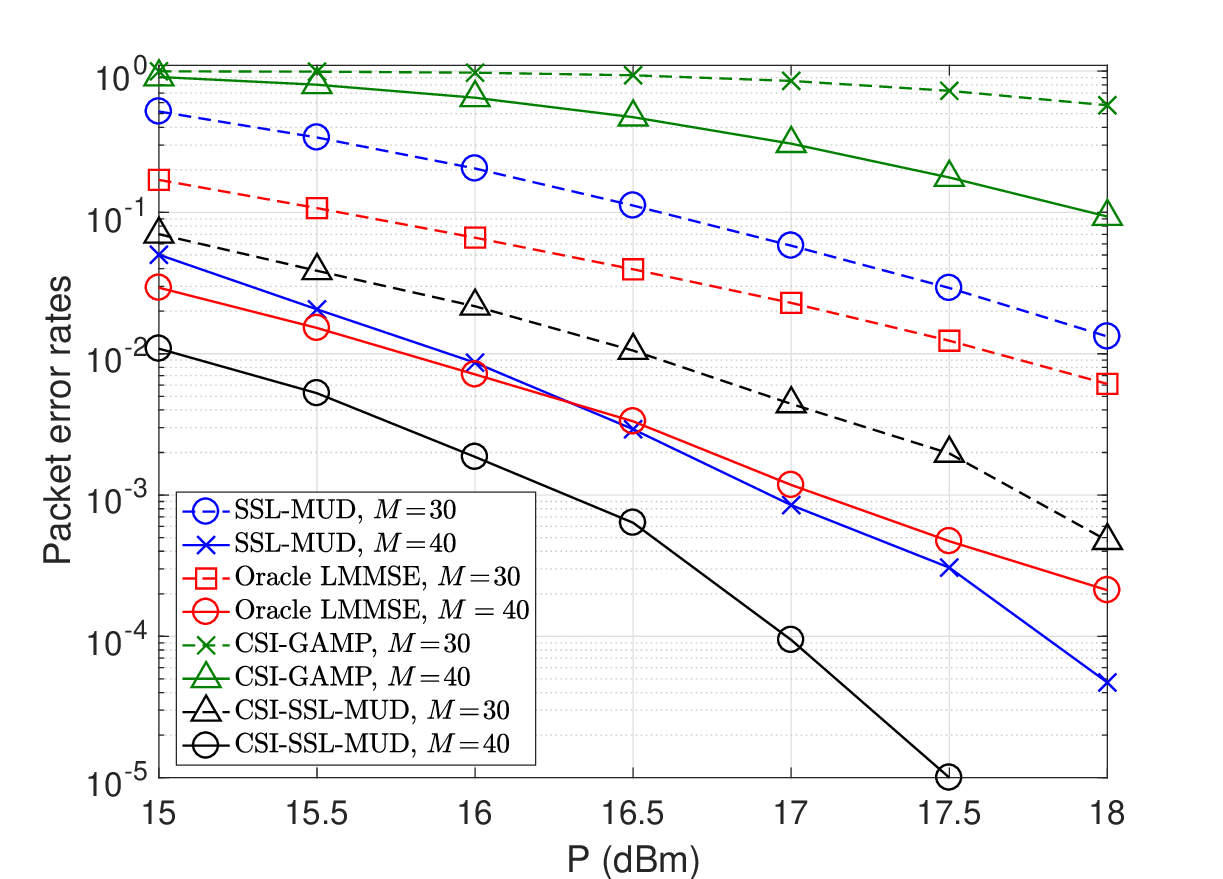}\\
	\caption{PER comparison between SSL-MUD and the three baseline schemes for system settings $\lambda = 1/2000$, $M = 30$ and $40$ respectively.}\label{fig_NTS_1}
\end{figure}

\begin{figure}
	\center
	\includegraphics[width=8cm]{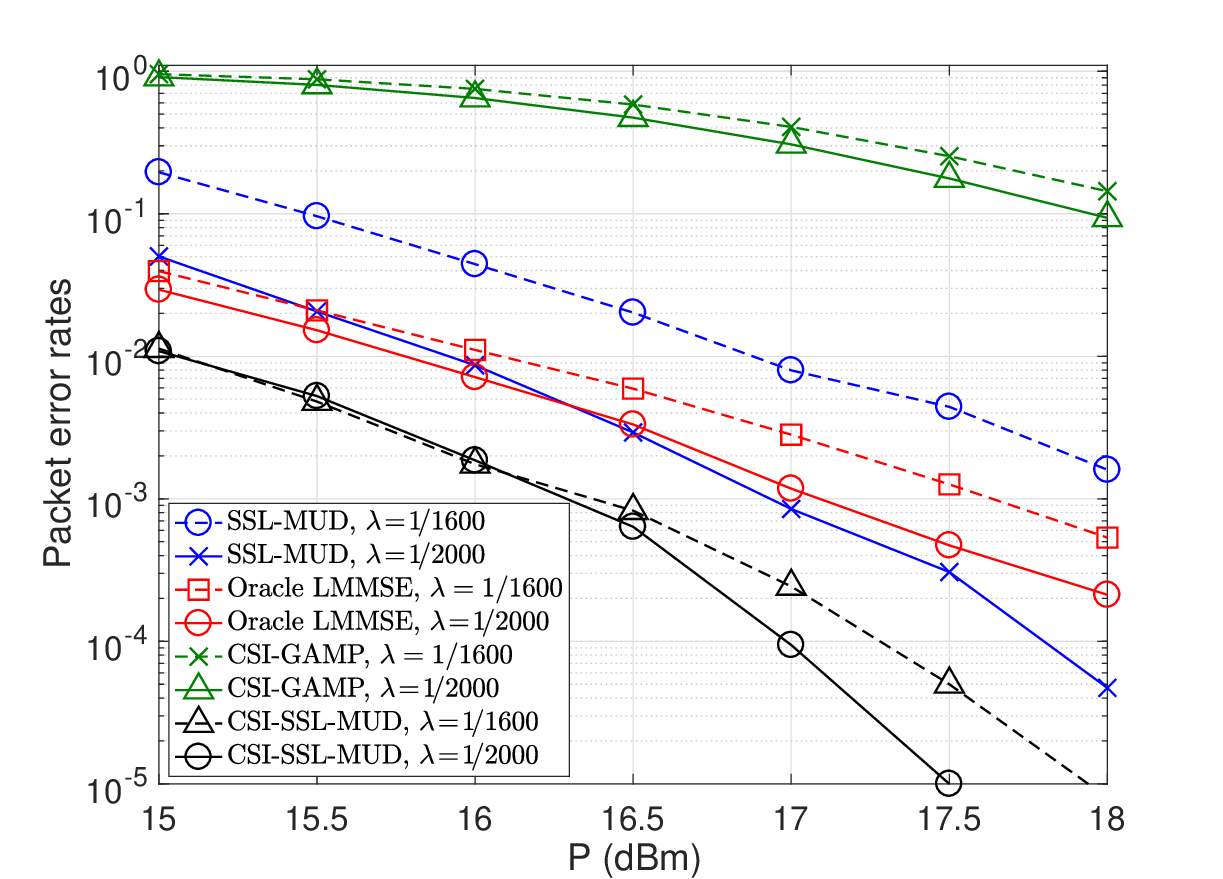}\\
	\caption{PER comparison between SSL-MUD and the three baseline schemes for system settings $M = 40$, $\lambda = 1/2000$ and $1/1600$, respectively. For $\lambda = 1/2000$, the average and the maximum values of $N$ are $30.8$ and $54$, respectively. For $\lambda = 1/1600$, the average and the maximum values of $N$ are $38.4$ and $66$, respectively.}\label{fig_NTS_2}
\end{figure}

\begin{figure}
	\center
	\includegraphics[width=8cm]{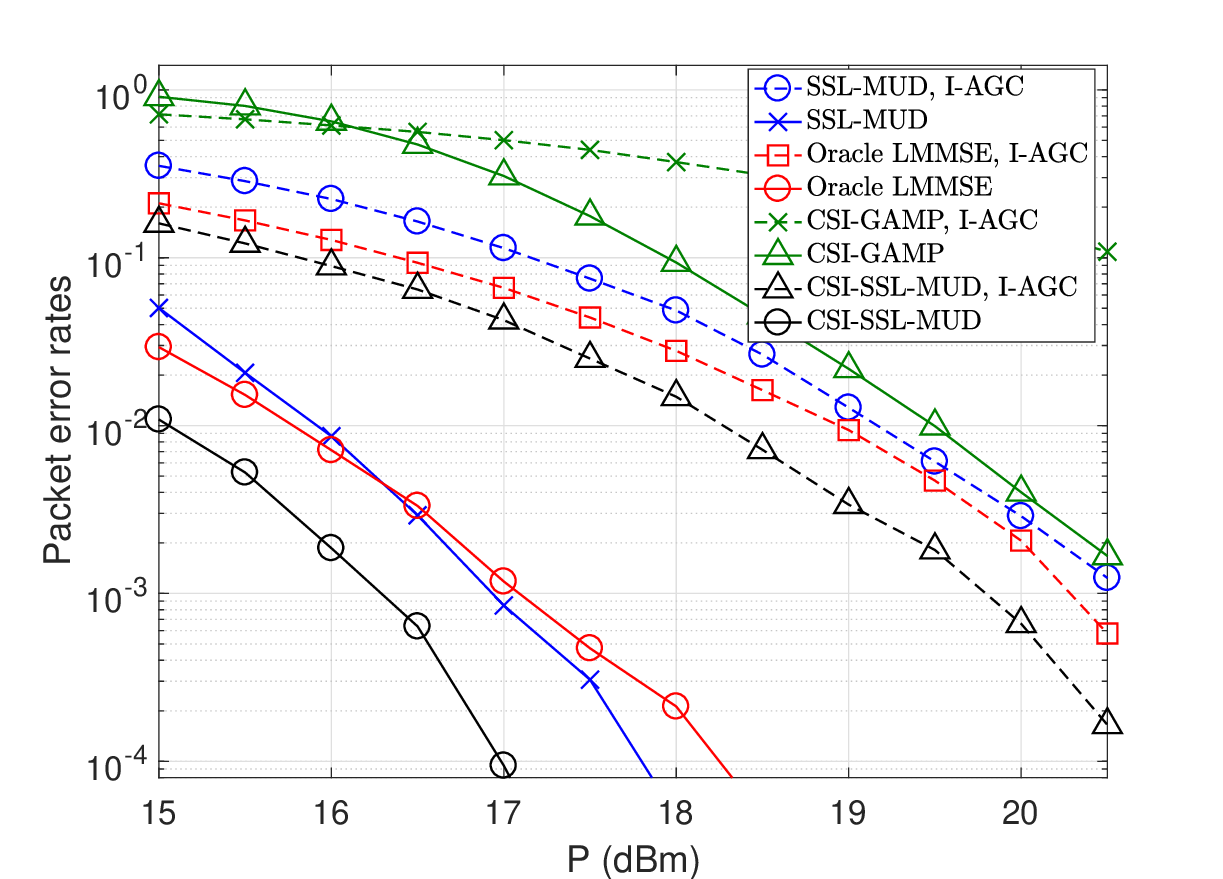}\\
	\caption{PER comparison between SSL-MUD and the three baseline schemes under perfect and imperfect AGC. The term ``I-AGC" refers to imperfect AGC.}\label{fig_NTS_3}
\end{figure}

The prior works mostly assume time-slotted transmission, and hence it is not convenient to compare their performance with that of SSL-MUD. For comparison, we introduce a heuristic CS based approach based on the ideas in \cite{ref_CSMUD_CDMA1, ref_CSMUD_CDMA2, ref_CSMUD_CDMA3, ref_CSMUD_CDMA4, ref_CSMUD_CDMA5, ref_CSMUD_CDMA6, ref_CSMUD_CDMA7}. In this heuristic approach, we assume that the BS perfectly knows the channel $\mathbf{H}$ in each observation window. Then the problem of estimating $\mathbf{X}$ from \eqref{eq_problem_NTS} can be solved by conventional CS approaches such as GAMP. In the simulations, we refer to this approach as `CSI-GAMP". We also include a comparison with a widely used benchmark in the prior works \cite{ref_CSMUD_CDMA2,ref_CSMUD_CDMA4,ref_CSMUD_CDMA3}, termed Oracle LMMSE, in which the LMMSE estimation is performed under the assumption of perfect knowledge of CSI and user activity state, i.e., the full matrix $\mathbf{H}$ and the locations of non-zero entries in $\mathbf{X}$. Note that Oracle LMMSE is more idealized than CSI-GAMP since the latter does not know the user activity state. A benchmark termed ``CSI-SSL-MUD" is also provided, where we run the Turbo-BiG-AMP algorithm with perfect knowledge of the channel. Therefore, the PER of CSI-SSL-MUD serves as a lower bound for the PER of SSL-MUD.

Fig. \ref{fig_NTS_1} investigates the PER performance of SSL-MUD and the three baseline schemes for $\lambda = 1/2000$, $M = 30$ and $40$, respectively. We see that SSL-MUD significantly outperforms CSI-GAMP, even though the latter assumes prior knowledge of CSI. This is because the conventional CS solvers such as GAMP assume that the sparse representation $\mathbf{X}$ consists of i.i.d. entries, while in the non-time-slotted grant-free MaDMA system the sparsity in $\mathbf{X}$ is highly structured. We also see that SSL-MUD achieves a PER comparable to that of Oracle LMMSE, and even outperforms Oracle LMMSE in the relatively high SNR regime for $M=40$. This implies that the structured sparsity in $\mathbf{X}$, if appropriately exploited, can be used to significantly reduce the overhead for the acquisition of user activity state and CSI. At the PER level of $10^{-2}$, SSL-MUD is 1dB and 1.5dB worse than CSI-SSL-MUD for $M=30$ and $40$, respectively, implying that SSL-MUD only suffers from a slight performance loss due to the absence of CSI. This demonstrates that as a blind detection scheme, SSL-MUD can significantly reduce the overhead for channel estimation without sacrificing much error performance.

Fig. \ref{fig_NTS_2} provides the PER comparison for $M = 40$, $\lambda = 1/2000$ and $1/1600$, respectively. Intuitively, a greater user activity rate reduces the sparsity in $\mathbf{X}$, increases the number of active packet $N$ in each observation window, and hence results in a worse PER performance. In the simulations, we observe that for $\lambda = 1/2000$, the average and the maximum values of $N$ are $30.8$ and $54$, respectively, while for $\lambda = 1/1600$, the average and the maximum values of $N$ are $38.4$ and $66$, respectively. Fig. \ref{fig_NTS_2} shows that under different user activity rates, SSL-MUD still achieves a performance comparable to that of Oracle LMMSE, and significantly outperforms CSI-GAMP.

In the above simulations, we assume perfect AGC so that the BS receives equal power from all active users. In practice, the AGC may be imperfect due to the inaccuracy of the amplifiers embedded in the user devices. To investigate the performance loss due to imperfect AGC, we run simulations where the receive powers of different users at the BS have a variation up to 5dB. More specifically, we randomly generate the value of each $\beta_u$ in dB from a uniform distribution over $[-130.5, -125.5]$. The simulation results are plotted in Fig. \ref{fig_NTS_3}. It is observed that at the PER level of $10^{-2}$, all schemes suffer from a loss of about 3dB due to the imperfection of AGC. Under both perfect and imperfect AGC, SSL-MUD significantly outperforms CSI-GAMP, and the performance gap between SSL-MUD and CSI-SSL-MUD is less than 1dB. This implies that the SSL-MUD scheme can tolerate unequal receive powers at the BS.

\section{Conclusion}
In this work, we developed two MUD schemes, namely RSL-MUD and SSL-MUD, for the time-slotted and non-time-slotted grant-free MaDMA systems, respectively. By exploiting the random sparsity (for time-slotted grant-free MaDMA) and the structured sparsity (for non-time-slotted grant-free MaDMA) of the user signals, the BS can recover the user activity state, the channel, and the user data through message-passing-based algorithms in a single phase without requiring pilot signals. Simulation results demonstrate that both RSL-MUD and SSL-MUD significantly outperform their counterpart schemes in the literature.

The study of grant-free MaDMA system is still in an initial stage. Based on our work, the following directions will be of interest for future research.
\begin{itemize}
	\item{\textbf{Channel coding}: So far, our sparsity learning based MUD scheme does not consider channel coding. Potentially, applying channel coding can significantly reduce the error rate of the packet recovery. Further, for channel codes such as the low-density parity-check (LDPC) codes, the decoding is also based on message passing. Therefore, the channel decoding process can be embedded in the BiG-AMP and Turbo-BiG-AMP algorithms to improve the detection efficiency.}
	\item{\textbf{Asynchronized multiple access}: In this work, we studied the non-time-slotted grant-free MaDMA system, where active users are required to be synchronized at the symbol level but not the packet level. In practice, symbol-level synchronization still introduces additional overhead. We believe that our SSL-MUD scheme can be extended to the asynchronized grant-free MaDMA system, where the clock difference between active users can be regarded as latent variables to be learned by the BS. The corresponding algorithm design is a challenging task worthy of future research endeavor.}
\end{itemize}

% if have a single appendix:
%\appendix[Proof of the Zonklar Equations]
% or
%\appendix  % for no appendix heading
% do not use \section anymore after \appendix, only \section*
% is possibly needed

% use appendices with more than one appendix
% then use \section to start each appendix
% you must declare a \section before using any
% \subsection or using \label (\appendices by itself
% starts a section numbered zero.)
%

%\appendices
%\section{Proof of the First Zonklar Equation}
%Appendix one text goes here.

% you can choose not to have a title for an appendix
% if you want by leaving the argument blank
%\section{}
%Appendix two text goes here.

% use section* for acknowledgment
%\section*{Acknowledgment}

%The authors would like to thank...

% Can use something like this to put references on a page
% by themselves when using endfloat and the captionsoff option.
\ifCLASSOPTIONcaptionsoff
  \newpage
\fi

\begin{IEEEbiography}[{\includegraphics[width=1in,height=1.25in,clip,keepaspectratio]{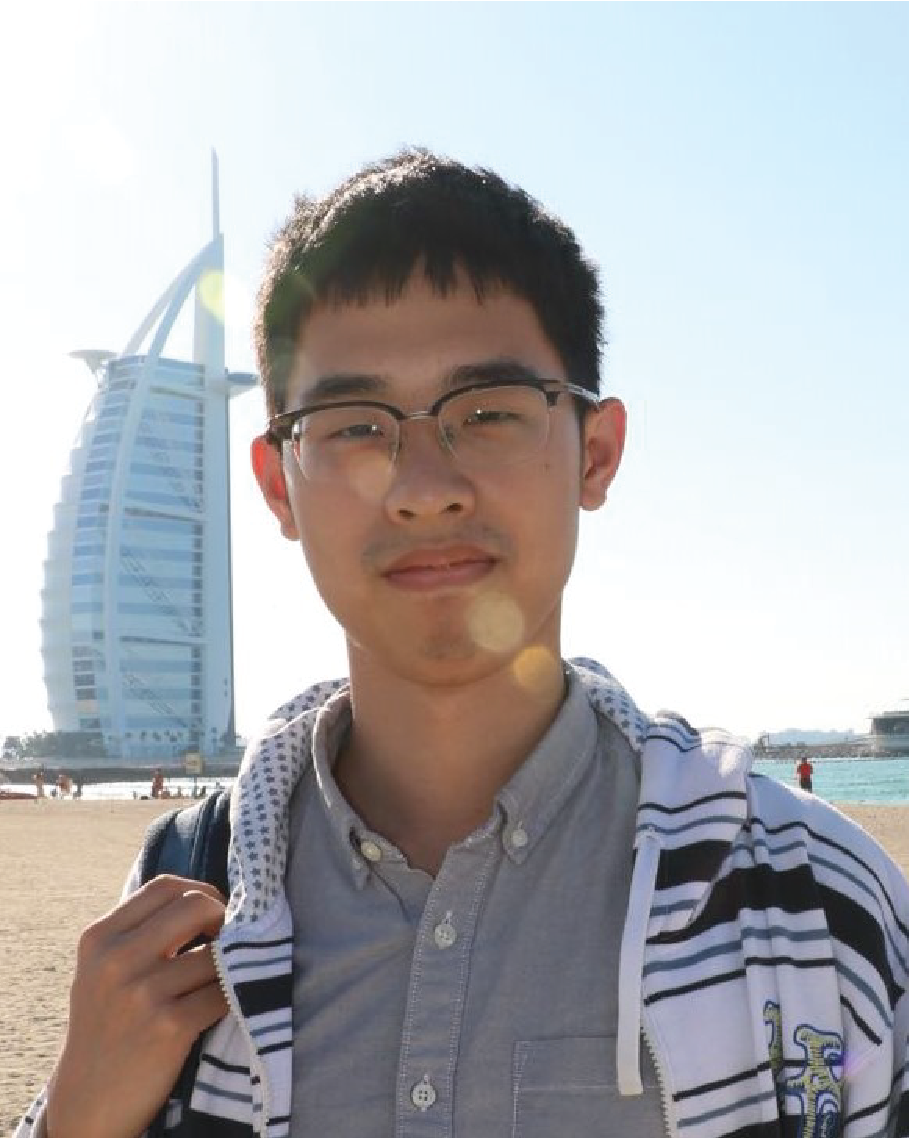}}]{Tian Ding}
received the B.Eng. degree in automation from Tsinghua University in 2014, and the M.Phil. degree in information engineering from The Chinese University of Hong Kong in 2016, where he is currently pursuing the Ph.D. degree with the Department of Information Engineering. His research interests include information theory, wireless networks, Internet of Things, and deep learning theory.
\end{IEEEbiography}

% if you will not have a photo at all:
\begin{IEEEbiography}[{\includegraphics[width=1in,height=1.25in,clip,keepaspectratio]{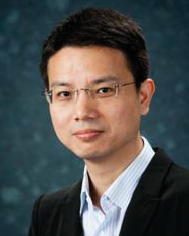}}]{Xiaojun Yuan} (S’04–M’09–SM’15) received the Ph.D. degree in electrical engineering from the City University of Hong Kong, in 2008. From 2009 to 2011, he was a Research Fellow of the Department of Electronic Engineering, City University of Hong Kong. He was also a Visiting	Scholar with the Department of Electrical Engineering, University of Hawaii at Manoa, in spring and summer 2009, and in 2010. From 2011 to	2014, he was a Research Assistant Professor with the Institute of Network Coding, The Chinese University of Hong Kong. From 2014 to 2017, he was an Assistant Professor with the School of Information Science and Technology, ShanghaiTech University. He is currently a Professor with the Center for Intelligent Networking and Communications, University of Electronic Science and Technology of China, supported by the Thousand Youth Talents Plan in China. 
	
His research interests include a broad range of signal processing, machine learning, and wireless communications, including but not limited to multi-antenna and cooperative communications, sparse and structured signal recovery, Bayesian approximate inference, and	network coding. He has published over 130 peer-reviewed research papers in the leading international journals and conferences in the related areas. He was a co-recipient of the Best Paper Award of the IEEE International Conference on Communications (ICC) 2014, and also a co-recipient of the Best Journal Paper Award of the IEEE Technical Committee on Green Communications and Computing (TCGCC) 2017. He has served on a number of technical programs for international conferences. He is an Editor of the IEEE Transactions on Communications, since 2017, and also an Editor of the IEEE Transactions on Wireless Communications, since 2018.

\end{IEEEbiography}

% insert where needed to balance the two columns on the last page with
% biographies
%\newpage

\begin{IEEEbiography}[{\includegraphics[width=1in,height=1.25in,clip,keepaspectratio]{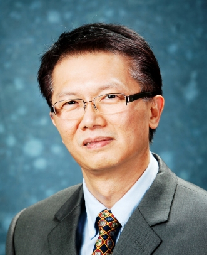}}]{Soung Chang Liew}(S'84-M'87-SM'92-F'12) received the S.B., S.M., E.E., and Ph.D. degrees from the Massachusetts Institute of Technology. From 1984 to 1988, he was with the MIT Laboratory for Information and Decision Systems, where he investigated fiber-optic communications networks. From 1988 to 1993, he was with Telcordia, NJ, where he involved in broadband network research. He has been a Professor with the Department of Information Engineering, The Chinese University of Hong Kong (CUHK), since 1993. He is currently the Division Head of the Department of Information Engineering and a Co-Director of the Institute of Network Coding, CUHK. He is also an Adjunct Professor with Peking University and Southeast University, China.
	
His research interests include wireless networks, Internet of Things, intelligent transport systems, Internet protocols, multimedia communications, and packet switch design. His research group received the best paper awards in the IEEE MASS 2004 and the IEEE WLN 2004. Separately, TCP Veno, a version of TCP to improve its performance over wireless networks proposed by Prof. Liew's Research Group, has been incorporated into a recent release of Linux OS. In addition, he initiated and built the first interuniversity ATM network testbed in Hong Kong in 1993. His research group pioneered and advanced the development of physical-layer network coding.
	
Besides academic activities, he is active in the industry. He co-founded two technology start-ups in Internet software and has been serving as a consultant to many companies and industrial organizations.
	
He holds 11 U.S. patents and a fellow of IET and HKIE. He was a recipient of the first Vice-Chancellor Exemplary Teaching Award in 2000 and the Research Excellence Award in 2013 at The Chinese University of Hong Kong.
\end{IEEEbiography}

% You can push biographies down or up by placing
% a \vfill before or after them. The appropriate
% use of \vfill depends on what kind of text is
% on the last page and whether or not the columns
% are being equalized.

%\vfill

% Can be used to pull up biographies so that the bottom of the last one
% is flush with the other column.
%\enlargethispage{-5in}

% that's all folks

\begin{thebibliography}{99}
	\bibitem{ref_WSN}
	M. A. Kafi, D. Djenouri, J. Ben-Othman, and N. Badache, ``Congestion control protocols in wireless sensor networks: A survey," \textit{IEEE Commun. Surveys \& Tutorials}, vol. 16, no. 3, pp. 1369-1390, 2014.
	
	\bibitem{ref_IoT}
	H. S. Dhillon, H. Huang, and H. Viswanathan, ``Wide-area wireless communication challenges for the internet of things," \textit{IEEE
		Commun. Mag.}, vol. 55, no. 2, pp. 168-174, Feb. 2017.
	
	\bibitem{ref_contention1}
	M. Hasan, E. Hossain, and D. Niyato, ``Random access for machine-to-machine communication in LTE-advanced networks: Issues and approaches,", \textit{IEEE Commun. Mag.}, vol. 51, no. 6, pp. 86-93, Jun. 2011.
	
	\bibitem{ref_contention2}
	N. K. Pratas, H. Thomsen, C. Stefanovic, and P. Popovski, ``Code-expanded random access for machine-type communications," in \textit{Proc. IEEE Globecom Workshops}, Dec. 2012.
	
	\bibitem{ref_contention3}
	E. Bj{\"o}ornson, et at., ``A random access protocol for pilot allocation in crowded massive MIMO systems," \textit{IEEE Trans. Wireless Commun.}, vol. 16, no. 4, pp. 2220-2234, Apr. 2017.
	
	\bibitem{ref_NOMA}
	L. L. Dai et al., ``Non-orthogonal multiple access for 5G: Solutions, challenges, opportunities, and future research trends," \textit{IEEE Commun. Mag.}, vol. 53, no. 9, pp. 74-81, Sept. 2015.
	
	\bibitem{ref_CS}
	D. L. Donoho, ``Compressed sensing," \textit{IEEE Trans. Inf. Theory}, vol. 52, no. 4, pp. 1289-1306, Apr. 2006.
	
	\bibitem{ref_model_CS}
	R. G. Baraniuk, V. Cevher, M. F. Duarte, and C. Hegde, ``Model-based compressive sensing," \textit{IEEE Trans. Inf. Theory, vol}. 56, no. 4, pp. 1982-2001, Apr. 2010.
	
	\bibitem{ref_CSMUD_CDMA1}
	H. Zhu and G. B. Giannakis, ``Exploiting sparse user activity in multiuser detection," \textit{IEEE Trans. Commun.}, vol. 59, no. 2, pp. 454-465, Feb. 2011.
	
	\bibitem{ref_CSMUD_CDMA2}
	H. F. Schepker, C. Bockelmann, and A. Dekorsy, ``Efficient detectors for joint compressed sensing detection and channel decoding," \textit{IEEE Trans. Commun.}, vol. 63, no. 6, pp. 2249-2260, Jun. 2015.
	%oracle LS
	
	\bibitem{ref_CSMUD_CDMA3}
	C. Wei et al., ``Approximate message passing based joint user activity and data detection for NOMA," \textit{IEEE Commun. Lett.}, vol. 21, no. 3, pp. 640-643, Mar. 2017. %oracle LMMSE
	
	\bibitem{ref_CSMUD_CDMA4}
	Y. Du et al., ``Efficient multi-user detection for uplink grant-free NOMA: Prior-information aided adaptive compressive sensing perspective," \textit{IEEE J. Sel. Areas
		Commun.}, vol. 35, no. 12, pp. 2812-2828, Jul. 2017. %oracle LS
	
	\bibitem{ref_CSMUD_CDMA5}
	X. Zhang, Y. Liang, and J. Fang, ``Bayesian learning based multiuser detection for M2M communications with time-varying user activities," in \textit{IEEE Int. Conf. Commun.
		(ICC)}, Paris, France, 2017.
	
	\bibitem{ref_CSMUD_CDMA6}
	B. Wang, L. Dai, T. Mir, and J. Li, ``Dynamic compressive sensing based multi-user detection for uplink grant-free NOMA," \textit{IEEE Commun. Lett.}, vol. 20, no. 11, pp. 2320-2323, Nov. 2016.
	
	\bibitem{ref_CSMUD_CDMA7}
	B. Wang, L. Dai, T. Mir, and Z. Wang, ``Joint user activity and data detection based on structured compressive sensing for NOMA," \textit{IEEE Commun. Lett.}, vol. 20, no. 7, pp. 1473-1476, Jul. 2016.
	
	\bibitem{ref_CSMUD_other1}
	T. Wang, et al., ``Block-sparse compressive sensing based multi-user and signal detection for generalized spatial modulation in NOMA," in \textit{IWCMC}, Valencia, Spain, 2017.
	
	\bibitem{ref_CSMUD_other2}
	Z. Gao, L. Dai, Z. Wang, S. Chen, and L. Hanzo, ``Compressive-sensing-based multiuser detector for the large-scale SM-MIMO uplink", \textit{IEEE
		Trans. Veh. Technol.}, vol. 65, no. 10, pp. 8725-8730, Oct. 2016.
	%oracle LS
	
	\bibitem{ref_CSMUD_other3}
	X. Meng, S. Wu, L. Kuang, D. Huang, and J. Lu, ``Multi-user detection for spatial modulation via structured approximate message passing," \textit{IEEE Commun. Lett.}, vol. 20, no. 8, pp. 1527-1530, Aug. 2016.
	
	\bibitem{ref_CSMUD_other4}
	T. Wang, S. Liu, F. Yang, J. Wang, J. Song and Z. Han, ``Generalized spatial modulation-based multi-user and signal detection scheme for terrestrial return channel with NOMA," \textit{IEEE Tran. Broadcast.}, vol. 64, no. 2, pp. 211-219, Jun. 2018.
	
	\bibitem{ref_CSMUD_noCSI1}
	G. Hannak, M. Mayer, A. Jung, G. Matz, and N. Goertz, ``Joint channel estimation and activity detection for multiuser communication systems," in \textit{IEEE Int. Conf. Commun. (ICC) Workshop}, Jun. 2015.
	
	\bibitem{ref_CSMUD_noCSI2}
	G. Wunder, P. Jung, and M. Ramadan, ``Compressive random access using a common overloaded control channel," in \textit{IEEE Int. Conf.
		Commun. (ICC) Workshops}, Jun. 2015.
	
	\bibitem{ref_CSMUD_noCSI3}
	X. Xu, X. Rao, and V. K. N. Lau, ``Active user detection and channel estimation in uplink CRAN systems," in \textit{IEEE Int. Conf. Commun.
		(ICC)}, pp. 2727-2732, London, UK, 2015.
	
	\bibitem{ref_CSMUD_noCSI4}
	L. Liu and W. Yu, ``Massive connectivity with massive MIMO--Part I: Device activity detection and channel estimation," \textit{IEEE Trans. Signal Process.}, vol. 66, no. 11, pp. 2933-2946, 2018.
	
	\bibitem{ref_CSMUD_noCSI5}
	L. Liu and W. Yu, ``Massive connectivity with massive MIMO--Part II: Achievable rate characterization," \textit{IEEE Trans. Signal Process.}, vol. 6, no. 11, pp. 2947-2959, Jun. 1, 2018.
	
	\bibitem{ref_CSMUD_noCSI6}
	Z. Chen, F. Sohrabi, and W. Yu, ``Sparse activity detection for massive connectivity," \textit{IEEE Trans. Signal Process.}, vol. 66, no. 7, pp. 1890-1904, Apr. 2018.
	
	\bibitem{ref_CSMUD_noCSI7}
	Y. Du et al., ``Joint channel estimation and multiuser detection for uplink grant-free NOMA," to appear in \textit{IEEE Commun. Lett.}, Feb. 2018.
	
	\bibitem{ref_sparse_bayesian1}
	M. Tipping, ``Sparse Bayesian learning and the relevance vector machine," \textit{J. Mach. Learning Res.}, vol. 1, pp. 211-244, 2001.
	
	\bibitem{ref_sparse_bayesian2}
	R. Prasad, C. Murthy, and B. Rao, ``Joint channel estimation and data detection in MIMO-OFDM systems: A sparse Bayesian learning approach," \textit{IEEE Trans. Signal Process.}, vol. 63, no. 20, pp. 5369-5382, Oct 2015.
	
	\bibitem{ref_sparse_bayesian3}
	S. Liu, F. Yang, J. Song, and Z. Han, ``Block sparse Bayesian learning based NB-IoT interference elimination in LTE-Advanced systems,", \textit{IEEE Trans. Commun.}, vol. 65, no. 10, pp. 4559-4571, October 2017.
	
	\bibitem{ref_dictionary_learning}
	K. Kreutz-Delgado et al., ``Dictionary learning algorithms for sparse representation," \textit{Neural Comput.}, vol. 15, no. 2, pp. 349-396, 2003.
	
	\bibitem{ref_BiGAMP}
	J. T. Parker, P. Schniter, and V. Cevher, ``Bilinear generalized approximate message passing Part I: Derivation," \textit{IEEE Trans. Signal Process.}, vol. 62, no. 22, pp. 5839-5853, Nov. 2014.
	
	\bibitem{ref_Turbo1}
	J. Ma, X. Yuan, and L. Ping, ``Turbo compressed sensing with partial DFT
	sensing matrix," \textit{IEEE Signal Process. Lett.}, vol. 22, no. 2, pp. 158-161, Feb. 2015.
	
	\bibitem{ref_Turbo2}
	Z. Xue, J. Ma, and X. Yuan, ``Denoising-based turbo compressed sensing," \textit{IEEE Access}, vol. 5, pp. 7193-7204, Apr. 2017.
	
	\bibitem{ref_Turbo3}
	J. Vila, P. Schniter, and J. Meola. ``Hyperspectral unmixing via turbo bilinear generalized approximate message passing," \textit{IEEE Trans. Comput. Imag.}, vol. 21, no.3, pp. 143-158, Aug. 2015.
	
	\bibitem{ref_SCMA}
	H. Nikopour and H. Baligh, ``Sparse Code Multiple Access," in \textit{Proc. IEEE PIMRC}, London, UK, Sep. 2013, .
	
	\bibitem{ref_Onsager}
	D. L. Donoho, A. Maleki, and A. Montanari, ``Message-passing algorithms for compressed sensing," \textit{Proceedings of the National Academy of Sciences of the United States of America,} vol. 106, no. 45, pp. 18914-18919, 2010.
	
	\bibitem{ref_blind}
	J. Zhang, X. Yuan, and Y. J. Zhang, ``Blind signal detection in massive MIMO: Exploiting the channel sparsity," \textit{IEEE Trans. Commun.,} vol. 66, no. 2, pp. 700-712, Feb. 2018.
	
	\bibitem{ref_GAMP}
	S. Rangan, ``Generalized approximate message passing for estimation with random linear mixing," in \textit{IEEE International Symposium on Information Theory (ISIT)}, pp. 2168-2172, Saint-Petersburg, Russia, 2011.
	
	\bibitem{ref_sphDec}
	B. Hassibi and H. Vikalo, ``On the sphere-decoding algorithm I. Expected complexity," \textit{IEEE Trans. Signal Process.}, vol. 53, no. 8, pp. 2806-2818, Jul. 2005.
	
	\bibitem{ref_Conf}
	T. Ding, X. Yuan, and S. C. Liew, ``Structured sparsity learning based multiuser detection in massive-device multiple access", in \textit{Proc. IEEE Global Communications Conference (GLOBECOM)}, Abu Dhabi, UAE, Dec. 2018.
	
\end{thebibliography}
\end{document}